\documentclass{article}
\usepackage{cite} 
\usepackage{url}  
\usepackage{ifthen}  
\usepackage{multicol}   
\usepackage[utf8]{inputenc} 
\usepackage{amsmath, amssymb, amsfonts, upgreek}
\usepackage{tabularx}
\usepackage{graphicx}
\usepackage{multirow}
\usepackage{booktabs}
\usepackage{caption}

\begin{document}

\title{Generalization of the normal-exponential model: exploration of a more accurate parametrisation for the signal distribution on Illumina BeadArrays}

 \author{Sandra Plancade\,$^{1}$\footnote{to whom correspondence should be addressed}, Yves Rozenholc\,$^{2}$, Eiliv Lund\,$^{1}$}
 \date{}

\maketitle

\noindent $^{1}${\small Department of Community Medicine, Faculty of Health Sciences,
University of Troms\o{}, 9037 Troms\o{}, Norway.}\\
$^{2}${\small Department of Applied Mathematics, MAP5, 45 rue des Saints-P\`eres, University Paris Descartes, 75006 Paris.}

\begin{abstract}
        
\textbf{Motivation:}
Illumina BeadArray technology includes non specific negative control features that allow a precise estimation of the background noise. 
As an alternative to the background subtraction proposed in BeadStudio which leads to an important loss of information by generating negative values, a background correction method modeling the observed intensities as the sum of the exponentially distributed signal and normally distributed noise has been developed. Nevertheless, Wang and Ye \cite{Wang11} display a kernel-based estimator of the signal distribution on Illumina BeadArrays and suggest that a gamma distribution would represent a better modeling of the signal density. Hence, the normal-exponential modeling may not be appropriate for Illumina data and background corrections derived from this model may lead to wrong estimation.  

\textbf{Results:}
We propose a more flexible modeling  based on a gamma distributed signal and a normal distributed background noise and develop the associated background correction. Our model proves to be markedly more accurate to model Illumina BeadArrays: on the one hand, it is shown on two types of Illumina BeadChips that this model offers a more correct fit of the observed intensities. On the other hand, the comparison of the operating characteristics of several background correction procedures on spike-in and on normal-gamma simulated data shows high similarities, reinforcing the validation of the normal-gamma modeling. 
The performance of the background corrections based on the normal-gamma and normal-exponential models are compared on two dilution data sets, through testing procedures which represent various experimental designs. Surprisingly, we observe that the implementation of a more accurate parametrisation in the model-based background correction does not increase the sensitivity. These results may be explained by the operating characteristics of the estimators: the normal-gamma background correction offers an improvement in terms of bias, but at the cost of a loss in precision. 

\textbf{Availability:}
The R-code to perform the background correction in this new model are available in the R package {\tt NormalGamma} (http://cran.r-project.org/).

\end{abstract}

\section{Background}\label{sec:intro}

Illumina BeadArray platform is a microarray technology offering highly replicable measurements of gene expression in a biological sample. Each probe is measured on average of thirty to sixty beads randomly distributed on the surface of the array, avoiding spatial artifacts and the reported probe intensity is the  robust mean of the bead measurements. Fluorescence intensity measured on each bead is subject to several sources of noise  (non-specific binding, optical noise, ...). Thus the intensities produced by the microarray require a background correction so that it takes into account measurement error.
For that purpose, Illumina microarray design includes   a set of non specific negative control probes which provides an estimate of the background noise distribution. 

In genome-wide microarrays, the  observed  intensity of a probe is usually modeled as  the sum of a signal and a background noise. Namely, let $X$ be the observed intensity of a given probe, we assume that
\begin{equation}\label{eq:additive-model} 
X= S +B
\end{equation}
where $S$ is the true signal which counts for the abundance of the probe complementary sequence in the target sample and is independent of the background noise $B$. Only $X$ is observed but the quantity of interest is the signal $S$. Therefore, a background correction adjusting the effect of noise on the true signal is necessary to enhance the biological validity of the results. In this context the knowledge of both signal and noise distributions provides a background correction procedure: the signal $S$ is estimated by the conditional expectation of $S$ given the observation $X=x$ and given the distributions of $B$ and $S$. Under parametric assumptions on $B$ and $S$, the problem is limited to the estimation of the parameters.
Besides, in many experimental contexts involving measurement error the normal distribution of the noise is assumed. Specific arguments for microarray data find their origins in analytical chemistry (see e.g. \cite{Rocke01}).

Background correction of  Affymetrix and two-color microarray data has been widely developed in literature (see \cite{Ritchie07} for a review and a comparison). Irizarry et al \cite{Irizarry03} proposed a parametric model for Affymetrix based on a exponential distribution of the signal, called normexp model. Several estimation procedures have been developed for this model. The first parameter estimation, still popular today, is the  Robust Multi-array Average (RMA) procedure. Maximum Likelihood Estimation (MLE),  incorporating the negative controls, has been later proposed and is considered to be more sensitive to the true parameter values (see \cite{Silver08}).
These procedures can be found in  Bioconductor\footnote{\tt http://www.bioconductor.org} packages including {\tt limma} \cite{Smyth05}. 

Illumina design differs from those of Affymetrix and two-color microarrays by  including a set of negative probes which do not specifically target any regular probe. Aside from non specific hybridization, these  negative probes  do not  hybridize and then have  signals close to zero. Thus their observed intensity is 
$ X=B.$ As all probes from a given array correspond to the same biological sample and are subject to the same technical steps during the analysis process, the noise is generally assumed identically distributed on an array and  the negative probes provide a  sample from its distribution. 

The background correction implemented in Illumina BeadStudio software is the subtraction  of the estimated mean of the negative probe distribution. However, it creates  a large amount of probes with negative intensities unusable in further analysis. The deletion of these probes is considered in some studies as an opportunity to gain statistical power when the number of strongly differentially expressed genes is large, but it can lead to an important loss of information. Ding et al \cite{Ding08} illustrate this phenomenon in their mice leukemia study:   a large amount of  corrected values  are negative only in one group suggesting that the corresponding probes have discriminating ability. This issue is confirmed by Dunning et al \cite{Dunning08} on  spike-in data.

To avoid this problem, parametric models have been used on Illumina data with parameter estimations taking into account the specific design of Illumina microarrays. 
 In this context, the normexp model has been first adapted.
 Ding et al \cite{Ding08} use the Maximum Likelihood Estimation (MLE) based on a Monte-Carlo Markov chain approximation and compare their method to an Illumina-adapted RMA procedure using an {\it ad hoc} rule of thumbs to estimate the parameters. Xie et al \cite{Xie09} go into details in normexp method comparison on experimental and simulated data. 
 Lin et al \cite{Lin08} present a variance stabilizing transformation (VST) on a model involving both additive and multiplicative noises, which simultaneously denoise and transform the data. Replacing the classical $\log$-transformation, VST produces  less directly interpretable results and tends to produces very small fold changes, as underlined by Shi et al \cite{Shi10} who propose an original approach to compare methods offering different bias-precision trade-off by aligning the innate offset generated by each pre-processing strategy. They conclude in favor of the normexp model with robust 'non-parametric' parameters associated to a quantile-normalization using control and regular probes.

The spread of each background correction among the Illumina users is hard to evaluate since many authors do not mention precisely the pre-processing steps performed in their study. Nevertheless, the normexp model that will be especially examined in this paper is included in several widely used packages such as {\tt lumi} and {\tt limma} of Bioconductor\footnotemark[1]. \medskip

Despite its popularity the normexp model does not properly fit Illumina microarray data. This issue has been raised by Wang and Ye \cite{Wang11}, who estimate the density of the signal on an Illumina microarray with a kernel-based deconvolution procedure. The shape of the estimated signal density does not present the characteristics of an exponential distribution and  a gamma modeling seems more appropriate. 
In this paper, we emphasize that the normal-exponential model is not flexible enough to model the signal-noise decomposition on Illumina microarrays by showing that the distance between the reconstructed density from the estimated parameters and the distribution of the observed intensities is large.

We propose an alternative model thereafter called  ``{\it normal-gamma} model" which addresses this lack of fit. In our model, the normal noise distribution is assumed and the signal on one array is assumed to be gamma distributed. As the exponential distribution is a special case of the gamma distribution, this model extends the normexp model. The potential of such generalization was already suggested by Xi ete al \cite{Xie09} in their discussion.
We derive the necessary estimation procedure by  likelihood maximization. The good quality of fit is attested  on two types of Illumina microarrays. The associated background correction is compared to methods based on the normexp model in terms of quality of estimation of the signal and checked for robustness on simulated data. The characteristics of the background correction procedures are compared on a set of spike-in data, and a parallel is drawn with the same characteristics studied on normal-gamma simulated data. Finally, the normexp and normal-gamma background corrections are compared on two dilution data sets.

The paper is organized as follows. The experimental and simulated data sets are described in Section \ref{sec:materials}. The methods are presented in Section \ref{sec:methods}: the notations and the general model-based background correction formula are gathered in Section \ref{sec:notations}; the previous models developed for Illumina microarray background correction, including the normexp model, are summarized in Section \ref{sec:normexp}; 
Section \ref{sec:normgam} presents the proposed alternative parametric model built with normal noise and gamma distributed signal, as well as a parametric estimation procedure and its associated background correction. In Section \ref{sec:evaluation}, the performances of this new model are evaluated on simulated, spike-in and dilution data sets. The results and perspectives are discussed in Section \ref{sec:discussion}. The tables and figures are gathered at the end of the article. Supplementary material is provided in Appendix. 
 
\section{Materials} \label{sec:materials}
\medskip

\subsection{Experimental Data sets}

\begin{itemize}

\item $(E_1)$ \textbf{Nowac data} \cite{Lund-nowac}. The gene expression profile in peripheral blood from ten controls in the Norwegian Woman And Cancer study has been analysed on Human HT-6 v4 Expression BeadChips. The whole probe set including 48,000 bead types has been considered, as well as a restricted set of 25,000 bead types according to Illumina annotation files. Details on laboratory experiments are given in Supplementary Material (SM), Section 1.1.

\item $(E_2)$ \textbf{ Leukemia mice data} \cite{Ding08}.
Total RNA from samples of spleen cells from four mice have been analysed on Mouse-6 v1 BeadChips. Experiment description and data are available in \cite{Ding08}

\item $(E_3)$ \textbf{ Spike-in data} \cite{Irizarry08}.   HumanWG-6 v2 BeadChips have been  customized to include 34 bead types, refered as 'spikes', whose corresponding target sequence is absent from the human genome in addition to the 48,000 regular probes. The 34 spikes were introduced at 12 different concentrations (0pm, 0.01pm, 0.03pm, 0.1pm, 0.3pm, 1pm, 3pm, 10pm, 30pm, 100pm, 300pm, 1000pm) in a human biological sample. Each sample corresponding to a spike concentration has been analysed on four arrays.  The data are available at  http://rafalab.jhsph.edu/.

\item  $(E_4)$  \textbf{MAQC data} \cite{Shi-maqc}. Two pure samples, Universal Reference RNA (HBRR) and Human Brain Reference RNA (UHRR) were mixed in four different proportions (100\%/0\%, 75\%/25\%, 25\%/75\%, 0\%/100\%). Five replications of each sample have been analysed on HumanWG-6 v1 BeadChips. The data are available on GEO (access number GSE5350).

\item $(E_5)$ \textbf{Dilution data} \cite{Lynch10}. The pure samples UHRR and HBRR were mixed at different proportions (100\%/0\%, 99\%/1\%, 95\%/5\%, 90\%/10\%, 75\%/25\%, 50\%/50\%, 25\%/75\%, 10\%/90\%, 0\%/100\%). Each mixed sample has been analysed with four different starting RNA quantities (250ng, 100ng, 50ng, 10ng). Six HumanWG-6 v3 BeadChips were used. 

\end{itemize}
\subsection{Simulated data sets}

For each data set,  $N=100$  random arrays including a vector 
${\bf X}^{\ell}$ of length
 $n_{\text{reg}}=25000$ corresponding to the regular probe intensities  and a vector  $\textbf{X}^{0,\ell}$ of length $n_{\text{neg}}=1000$ corresponding to the negative probe intensities are generated. The values of the nine parameter sets as well as the details of the simulations are given in SM, Section 1.2.

\begin{itemize}

\item $(S_1)$ \textbf{Normal-gamma and normexp models}. For each repetition $\ell = 1, \dots , N$,  $\textbf{X}^{\ell}$ is generated as the sum of a gamma and a normal-distributed sample, and  $\textbf{X}^{0,\ell}$  is drawn from a normal distribution.
Six sets of parameters are computed from  two microarrays in data sets $(E_1)$ and $(E_2)$, based on normexp and normal-gamma models in order to get realistic values (sets 1-6). The normexp parameters are actually degenerated normal-gamma parameters where the shape is equal to 1. 

\item $(S_2)$ \textbf{Mixture noise distribution}. A mixture of normal and $\upchi ^2$ distributions with different proportions (0, 0.1, 0.25, 0.5, 0.75,1) is considered for the background noise. These distributions model a departure from normality with a heavier right tail for larger values of $p$. The mixture densities are presented in SM, Section 7.4. The signal is generated from a gamma distribution with parameter values from set 1. 

\item $(S_3)$ \textbf{Replicates}. We mimic replicate measurements from a biological sample by simulating $N$ arrays with the same signal sample generated from a gamma distribution. The background noise and negative probe intensities  are independently drawn from a normal distribution for each array.  The values of the parameters are computed from the first array in $(E_3)$ (set 7). Replicates from the normal-exponential model are drawn in the same way with parameter values estimated on the same array with two normexp estimates (sets 8 and 9). 

\item $(S_4)$ \textbf{Replicates with empirical background noise}. Similarly to $(S_3)$, the signal drawn from a gamma distribution with parameter values from set 7  is identical on each array. In order to get a realistic noise distribution,  the negative probe and background noise intensities are sampled from the global set of quantile-normalised negative probe intensities measured in the experimental data set $(E_3)$.

\end{itemize}

\section{Methods}\label{sec:methods}

\medskip

\subsection{General model-based background correction formula} \label{sec:notations}

\medskip

\subsubsection{Notations.}
Throughout this article, the background correction is processed on one single array corresponding to one biological sample. For a given probe $j$, we denote by $X_j$ the observed intensity, $S_j$ the non-observable underlying signal and $B_j$ its background noise. For a negative control probe, $S_j$ is assumed to be 0. Let  $J$ and $J_0$ be respectively the index of regular and negative probes on the array. We denote by $f_X$, $f_S$ and $f_B$ the densities of respectively the observed intensity, the unknown signal of interest and the background noise.  \\

We denote by $f ^{\text{norm}} _{\mu, \sigma^2} $ the density of the normal distribution with mean $\mu$ and variance $\sigma^2$ 
and by $ \phi $ and $\Phi$ the density and cumulative distribution function of the normal distribution with mean 0 and variance 1. We denote by $f ^{\text{exp}} _{\alpha }  = (1/\alpha) \exp ( -x/\alpha)$ the density of the exponential distribution with mean $\alpha$
and $f ^{\text{gam}} _{\theta ,k }  =  x^{k-1} \exp \left(-x/\theta \right) /( \theta ^k \Gamma (k) )$ the density of the gamma distribution with scale parameter $\theta$ and shape parameter $k$.
The exponential distribution is a special case with $k=1$ and $\theta=\alpha$.

Given a parametric density and a  procedure of estimation of its parameters, we call {\it plug-in} density, this density for the estimated parameters.

\subsubsection{Model-based background correction}\label{sub:BgC}

The  model based background correction (BgC) 
incorporates information from both signal and noise distributions. 
Under the additive model (\ref{eq:additive-model}) assuming independence of $S$ and $B$, $f_X$  is  the convolution product of $f_S$ and $f_B$. For an observed probe intensity $x$, the signal $S$ is estimated by the conditional expectation of $S$ given the observation $X=x$ and the densities $f_B$ and $f_S$ (more details can be found in \cite{Xie09}):
\begin{eqnarray} \nonumber\label{shat}\label{Shat-eq}
\widehat{S}(x) &= &\mathbb{E} \left[ S | X=x, f_B , f_S \right] = \int s \frac{ f_S(s) f_B(x-s)}{f_X(x)} ds \\
 &= &  \int s f_S (s) f_B(x-s) ds\Big/ \int  f_S (s) f_B(x-s) ds. \quad 
 \end{eqnarray}

\subsection{Previous modelings} \label{sec:normexp}
\medskip

\subsubsection{The normal model for negative probes}
The design of Illumina BeadArrays provides a sample of the background distribution through the negative probes. We have compared the density histogram of negative probes to the {\it plug-in} normal density $f^{\text{norm}}_{\hat\mu,\hat\sigma}$ obtained by using robust estimators of the parameters  on data sets $(E_1)$ and $(E_2)$.

The results of this comparison are presented in SM, Section 7.2.1.
As expected, the empirical distribution is essentially normal but we notice a slightly heavier right tail. It may be interpreted as intensities of wrongly designed negative probes which partially hybridize  with some material present in the biological sample.

\subsubsection{The normal-exponential model} The normexp model is a parametric model for the noise-signal decomposition on one array. We recall it briefly and refer for example to \cite{Xie09} for more details. For every probe  $j$, 
\begin{eqnarray*}
 && X_j = S_j + B_j, \\
 && S_j \sim \left\{\begin{array}{ll}  \text{Exp} (\alpha), &\text{ if } j\in J, \\0 &\text{ if }j\in J_0,\end{array}\right.\\
&&  B_j \sim \mathcal{N} (\mu, \sigma ^2),\\
&& S_j \perp B_j,
 \end{eqnarray*}
 where $\perp$ denotes the independence between variables. The parameters $( \mu , \sigma,\alpha )$ depend on the given array. 
 
For computational reason, the $X_j$'s are usually and often implicitly assumed to be independent.  
The existence of pathways between genes violates this assumption. Nevertheless, as a small proportion of genes are involved, results
are reliable.
According to the convolution structure (see Section \ref{sub:BgC}), the density of the $X_j$'s is:
\begin{equation}\label{dens-normexp}
f^{\text{nexp}}_{\mu, \sigma , \alpha}  (x) = \frac{1}{\alpha} \exp \left( \frac{\sigma ^2}{2\alpha^2} - \frac{x-\mu}{\alpha} \right)  \Phi \left( \bar x \right),
\end{equation}
where $\bar x=(x-\mu -\sigma^2/\alpha)/\sigma$.
Denoting  $\Theta=(\mu,\sigma,\alpha)$, from (\ref{shat}), the background corrected intensity for an observed intensity $x$ is:
\begin{equation}\label{shat-exp} \widehat S^{\text{nexp}}(x|\Theta) = \sigma\left(\bar x + \frac{\phi(\bar x)}{\Phi(\bar x)}\right).\end{equation}

\subsubsection{Normal-exponential model fit}
We consider the data sets $(E_1)$ and $(E_2)$. For each array, we apply the following procedure:

\begin{itemize}

\item Computation of the estimators $(\widehat{\mu}, \widehat{\sigma}, \widehat{\alpha})$ of the normexp model with the methods described in \cite{Xie09}:

\begin{enumerate}

\item Maximum Likelihood Estimation (MLE) using both regular and negative probes,

\item Robust Multiarray Analysis (RMA) adapted from Affymetrix method,

\item NP estimation obtained by the method of moments applied to negative and regular probes, 

\item Bayesian estimation. Note that the bayesian estimation results are not presented as they are nearly identical to MLE, as pointed out by Xie et al \cite{Xie09}. 
  
\end{enumerate}

\item For each parameter estimation method, plot of the {\it plug-in} density $f^{\text{nexp}} _ {\widehat{\mu}, \widehat{\sigma}, \widehat{\alpha}}$. 

\item Plot of an irregular density histogram of all regular probe intensities of the array using the R-package {\it histogram} available on the CRAN with default irregular setting (see \cite{rozenholc-thoralf}). Even though adaptive irregular histograms are not commonly used to describe microarray data, they have been proved to offer a better approximation in a general framework. Moreover, they appear especially relevant to estimate microarray distributions which present high irregularities. 

\end{itemize}

Figure \ref{Fig1} shows the results for this procedure on one array from $(E_1)$ after removal of imperfectly designed probes (more arrays are presented  
in SM, Section 7.2.2).
Apart from the RMA method, the estimated density does not fit the density histogram and even the RMA estimator is not satisfying from a statistical point of view. One can remark that RMA underestimates the high expressions while the other methods tend to overestimate their contributions.

Besides Xie et al \cite{Xie09} show that the MLE and the NP estimation provide satisfying estimators of the parameters on normexp simulated data. 
Thus the difference between the histogram of the observed intensities and the plug-in density does not come from a poor estimation of the parameters but results from an unsuitable parametric model.

\subsection{A new modeling: The normal-gamma model} \label{sec:normgam} 

The poor fitting quality of the normexp model shown above calls for a more suitable parametric model for Illumina BeadArrays. According to Section \ref{sec:normexp} the normal assumption for the negative probes appears relevant. We consider the gamma distribution as an extension of the exponential distribution to model the signal intensities. Besides, as a scale mixture of exponential distributions (see \cite{Gleser89}), the gamma distribution is a natural generalization which helps to take into account different probe hybridization behaviors  which could count for different exponential life times. This defines a more flexible parametric model called the {\it normal-gamma} model that we propose to apply to Illumina BeadArrays.

\subsubsection{The normal-gamma model}
The normal-gamma model is defined as follows. 
For every probe  $j$:
\begin{eqnarray} \label{gam-nor-mod}\nonumber
 && X_j = S_j + B_j, \\\nonumber
 && S_j \sim \left\{\begin{array}{ll}  \Gamma(\theta,k), &\text{ if } j\in J, \\0 &\text{ if }j\in J_0,\end{array}\right.\\\nonumber
&&  B_j \sim \mathcal{N} (\mu, \sigma ^2),\\
&& S_j \perp B_j.
 \end{eqnarray}
 The parameters $(\mu , \sigma, k,\theta ) $ depend on the given array. This model offers more flexibility than the normexp model but requires the estimation of one more parameter.
 
According to the convolution structure (see Section \ref{sub:BgC}), the density of $X_j$ is the convolution product of the densities of $S_j$ and $B_j$, namely:

\begin{equation}\label{dens-normgam-conv}
f^{\text{ng}}_{\mu, \sigma , k,\theta }  (x) = \int f^{\text{gam}}_{k,\theta } (t) f^{\text{norm}}_{\mu, \sigma } (x-t) dt.
\end{equation}
This density does not have any analytic expression as the normexp density (\ref{dens-normexp}). Nevertheless,  good and fast numerical approximations can be computed using the Fast Fourier Transform ({\tt fft}) and some tail approximations to ensure stability. Our implementation based on  {\tt fft}  is detailed in SM, Section 7.6. 

\subsubsection{Parameter estimation in the normal-gamma model}
The parameters  $(\mu, \sigma , k, \theta)$ of the normal-gamma distribution are estimated by the Maximum Likelihood Estimator (MLE):

\begin{equation}\label{param-normgam}
(\widehat{\mu}, \widehat{\sigma}, \widehat{k}, \widehat{\theta}) = \arg \max _{(\mu, \sigma , k, \theta)} L\left( (\mu, \sigma , k, \theta) | \textbf{X}, \textbf{X}^0 \right)
\end{equation}
where 
\begin{equation*}
L\left( (\mu, \sigma , k, \theta) |  \textbf{X}, \textbf{X}^0\right) = \prod _{j\in J} f^{\text{ng}}_{\mu, \sigma ,k, \theta }  (X_j) \cdot \prod _{j\in J_0} f^{\text{norm}}_{\mu, \sigma }  (X_j)
\end{equation*}
is the likelihood from the two sets of observations $ \textbf{X} = \{ X_j, j\in J\} $ and $ \textbf{X} ^0= \{ X_j, j\in J_0\} $ measured on regular and negative probes, respectively.
Thanks to the  {\tt fft}-based approximation of $f^{\text{ng}}_{\mu, \sigma , k,\theta }$
the maximum likelihood estimation can be numerically computed using classical minimization algorithms (see SM, Section 7.6).

\subsubsection{Background corrected intensity for the normal-gamma model}
Denoting now $\Theta=(\mu, \sigma , k, \theta)$, we derive from (\ref{shat}) the background corrected intensity for an observed intensity $x$:
\begin{eqnarray}\label{Shat-eq1}
\nonumber \widehat{S}^{\text{ng}}(x|\Theta) &=& \int s f^{\text{gam}} _{k,\theta } (s) f^{\text{norm}} _{\mu, \sigma } (x-s)\,ds \Big/{f^{\text{ng}}_{\mu, \sigma , k, \theta }  (x) } \\
	&=& k\theta { f^{\text{ng}}_{\mu, \sigma , k+1,\theta }  (x)}\Big/{f^{\text{ng}}_{\mu, \sigma ,k, \theta }  (x)}
 \end{eqnarray}
using the equality $s f^{\text{gam}} _{k,\theta} (s) = k \theta  f^{\text{gam}} _{k+1,\theta } (s)$ valid for every $s>0$. This formula allows {\tt fft}-based computations for the background correction.

\subsubsection{Inference of negative probes from Illumina detection p-values}
Most  publicly available data sets do not present the negative probe intensities. Nevertheless, for each regular probe, Illumina provides a detection p-value equal to the proportion of negative probes which have intensities greater than that probe on a given array. Following the idea from Shi et al \cite{Shi10b} we propose to infer the negative probe intensities from the detection p-values (see details in SM, Section 7.7.1). 
For the normexp and the normal-gamma models, the estimates of the parameters and reconstructed signals obtained with the true and inferred negative probe intensities  are compared on the ten arrays from $(E_1)$. 
We observe that the error resulting from inference of the negative probe is negligible, with a relative error of order $10^{-3}$ to $10^{-4}$ on parameter estimation and $10^{-4}$ to $10^{-5}$ on signal estimation (see SM, Section 7.7.2).

\section{Results}\label{sec:evaluation}

\medskip

\subsection{Fit on Illumina BeadArray data}\label{section-normgam}

Similarly to Section $\ref{sec:normexp}$, we compare the irregular density histogram of the regular probe intensities, the {\it plug-in} normexp densities using RMA, MLE, NP methods, and  the {\it plug-in} normal-gamma density using a Maximum Likelihood Estimate of $(\mu, \sigma , k, \theta)$ on the data sets  $(E_1)$ and $(E_2)$. The results, similar along the arrays, are illustrated in Figure \ref{Fig2} on one array from $(E_1)$ (more plots are presented in SM, Section 7.2.2). We do not add the MLE and NP {\it plug-in} density estimates for the normexp model which have already been  shown not to fit the data.

Thanks to the larger flexibility of the normal-gamma model, we observe that the distance between the MLE {\it plug-in} normal-gamma density and the histogram of the intensities is smaller than the corresponding distance using the normexp model with any estimation procedure. This graphical result is confirmed numerically using the $L_1$-distance between the histogram and the reconstructed density defined by
\begin{equation}\label{absolute-deviation}
\ell_1(\hat f,\hat h) = \int |\hat f(x) - \hat h(x)|\,dx,
\end{equation}
where $\hat f$ represents one {\it plug-in} density estimate using either the normexp model or the normal-gamma model and  $\hat h$ represents the irregular density histogram obtained by using the R-package {\tt histogram}. Table \ref{Tab1}  presents the mean of the relative deviation for the normexp estimators with respect to the deviation for the normal-gamma estimator:
\begin{equation*}
 \text{mean} \left( \frac{ \ell_1(\hat f _i,\hat h_i)  }{ \ell_1(\hat f ^{\text ng}_i ,\hat h_i) } \right) 
 \end{equation*}
 where $\hat f_i$ is a normexp estimator of the regular probes density, and $\hat f ^{\text ng} _i $ is the normal-gamma estimator for individual $i$. The mean is computed over the ten arrays from $(E_1)$ (with and without the non specific binding probes) and over the four arrays from $(E_2)$.

The mean absolute deviation is 3 times smaller in favor of the normal-gamma density with respect to the normexp density using the RMA estimate, and 4 to 8 with respect to the normexp model using the MLE or NP estimates.

\subsection{Quality of estimation on simulated data}

The quality of estimation of the normal-gamma model is assessed on simulation the data set $(S_1)$.  The first two sets of parameters are non degenerate normal-gamma parameters, more realistic for modeling Illumina microarrays as shown in Section $\ref{section-normgam}$. They are used to evaluate the MLE normal-gamma parameter estimation and validate the associated background correction, and to quantify the improvement brought by the new normal-gamma background correction.  The last four sets are actually degenerate normal-gamma parameters where the shape parameter $k$ is set to 1, corresponding to normexp data, which enables to assess the potential loss of precision in parameter estimation brought by a more flexible modeling.

\subsubsection{Parameter estimation} 
 For each repetition $\ell = 1, \dots, N$ we compute the normal-gamma MLE $(\widehat{\mu}^{\ell}, \widehat{\sigma}^{\ell} ,\widehat{k}^{\ell}, \widehat{\theta}^{\ell})$ of the parameters.  Table \ref{Tab2}  presents the relative $L_1$-error for each parameter $\beta \in  \{\mu , \sigma,k,\theta \}$:
\begin{equation*}
\frac{1}{N  }\sum _{\ell =1}^N \left| \beta - \widehat{\beta}^{\ell}   \right| \Big/\beta.
\end{equation*}
The parameter estimation is of excellent quality for the gaussian distribution and of good quality for the gamma distribution.

To check wether the introduction of a fourth parameter in our model leads to a loss of precision  in the parameter estimation, we compare the relative $L_1$ errors of the MLE parameter estimation in the normal-gamma and normexp models using the parameter sets 3 to 6, corresponding to normexp data. The results summarized in Table \ref{Tab3} indicate that the quality is unchanged for the variance parameter and that we pay a price of order 2 for $\mu$ and $\theta$. Nevertheless,  since the relative errors in these cases have order $10^{-2}$, this loss is negligible.

\subsubsection{Background corrected intensity}\label{BgC}
We now study the performance of the normal-gamma background correction (BgC) obtained in (\ref{Shat-eq1}) with respect to the existing BgC methods (see \cite{Xie09}) in terms of quality of estimation of the signal on the simulated data set $(S_1)$. We compare the following BgC methods:
\begin{enumerate}\setcounter{enumi}{-1}
\item Normal-gamma BgC in (\ref{Shat-eq1}) with true parameters,

\item Normal-gamma BgC in (\ref{Shat-eq1}) with MLE parameters,

\item Normal-exponential BgC in (\ref{shat-exp}) with MLE parameters (referred to as normexp-MLE),

\item Normal-exponential BgC in (\ref{shat-exp}) with RMA parameters (referred to as normexp-RMA), 

\item Normal-exponential BgC in (\ref{shat-exp}) with NP parameters (referred to as normexp-NP),

\item Background subtraction: 
$$\widehat S^{\text{sub}}(x) = \max \big( x - \text{median} \{ X_j , j\in J_0 \} , 0 \big).$$

\end{enumerate}
These methods are further denoted by $\widehat S^{(i)}$ for $i=0,\ldots,5$. For methods 1 to 4, the BgC is a two-step procedure: the parameters are estimated and then plugged respectively into (\ref{Shat-eq1}) for method 1, and into (\ref{shat-exp}) for methods 2 to 4. From a practical point of view, as the parameters are unknown,  $\widehat S^{(0)}$ is unavailable. Nevertheless, as the result of a procedure with a perfect first estimation step, it allows a comparison to quantify the performance of the second step.

For each parameter set and for each BgC method $\widehat S=\widehat S^{(i)}$, $i=0,\ldots,5$ we compute the Mean Absolute Deviation (MAD):
 \begin{equation*}
\mathrm{MAD}(\widehat S) = \frac{1}{N} \sum _{\ell =1}^N \left( \frac{1}{n_{\text{reg}}} \sum _{j=1}^{n_{\text{reg}}} \left| \widehat{S}(X_j^\ell|\widehat\Theta_\ell) - S_j^\ell \right| \right)  
 \end{equation*}
where $\widehat \Theta_\ell=(\hat\mu_\ell,\hat\sigma_\ell,\hat k_\ell,\hat\theta_\ell)$ denotes for each simulated array $\ell$ the estimated parameters corresponding to the used methods with the following conventions: 1/ $\widehat\Theta_\ell$ is the true parameters for $i=0$; 2/ $\hat k_\ell=1$ for $i=2,\ldots,4$ corresponding to the exponential distribution; 3/ $\widehat\Theta_\ell$ represents the median over $\textbf{X}^{0,\ell}$ for $i=5$.  
Simulation results are summarized in Table \ref{Tab4}  for the parameter sets 1-6. The five first columns correspond to the excess risk ratio: 
\begin{equation}\label{mad-ratio}
R(i) = \mathrm{MAD}(\widehat S^{(i)})/\mathrm{MAD}(\widehat S^{(0)}), \quad \text{for } i=1,\ldots,5
\end{equation} 
and the last column indicates the reference risk $\mathrm{MAD}(\widehat S^{(0)})$.

The normal-gamma BgC provides the same quality when the parameters are known or estimated. This holds when the data are generated either from a normal-gamma or a normexp model. Normexp-NP shows good behaviors when the data come from a normexp model but has a risk increase of order 60\% if the data come from a normal-gamma model. Normexp-MLE provides good results for normal-exponential data but fails when the data come from a normal-gamma model. Not surprisingly, as already pointed by Xie et al \cite{Xie09}, normexp-RMA has a poor behavior. The background subtraction method with a maximal quality loss of 32\% offers an acceptable alternative in terms of risk. Indeed, most of the intensities being small, putting them to 0 does not affect significantly the MAD value. Let us recall, however, that  from a practical point of view, the main default of that BgC is to eliminate a huge number of probes. 

In practical experiments, the data are usually transformed before the analysis. To address this issue, the MAD is computed on log-transformed intensities (see SM, Section 7.3 for details), and the excess risk ratio is displayed in Table \ref{Tab5}. The normal-gamma BgC presents a smaller error of estimation than the methods based on the normexp model. The MAD from normexp-MLE generates the highest value, and normexp-RMA and normexp-NP show a a similar moderate excess risk. Nevertheless the differences between the BgC methods are less pronounced than the one observed at the raw scale. However, with an excess risk ratio between 16\% and 33\%, the normexp methods notably under perform the BgC based on the normal-gamma model in terms of signal estimation. \\

The MAD computation offers a global comparison of the various BgC methods in terms of signal estimation.  We refine this analysis by examining the absolute deviation (AD) of the estimated signal for each signal intensity at the raw and the log scales, respectively defined as:
$$ AD(S_j) = \frac{1}{N}  \sum _{\ell =1}^N  \left| \widehat{S}(X_j^{\ell} | \widehat{\Theta }_{\ell}) - S_j^{\ell} \right| $$
$$ AD(\log(S_j)) = \frac{1}{N}  \sum _{\ell =1}^N  \left| \log\left( \widehat{S}(X_j^{\ell} | \widehat{\Theta }_{\ell}) \right) - \log \left( S_j^{\ell} \right) \right| $$
 
The first row of Figure \ref{Fig3} displays the logarithm of the AD at the raw scale, as a function of the log-signal intensity. We observe that normexp-MLE presents a larger AD for all values of the signal. On small intensities, the normal-gamma BgC outperforms the other methods, whereas normexp-RMA and normexp-NP present a smaller deviation on moderate intensities. For high values of the signal, normal-gamma shows a smaller error of estimation together with normexp-NP.

The absolute deviation on log-transformed intensities is presented on the second row of Figure \ref{Fig3}. The normal-gamma BgC still presents the smallest error of estimation on weak intensities, but is outperformed by the other methods on moderate intensities. The four methods present similar AD values on high intensities. Besides, we observe than the error of estimation for all methods increases as the signal become weaker.

\subsubsection{Robustness}
In Section $\ref{sec:normexp}$, we have underlined the slightly heavier right tail of the negative probe distribution. To ensure that the estimation remains acceptable under the assumption of an imperfect noise parametrisation, we compare the robustness of the normal-gamma method with normexp-NP, stated as the most robust by \cite{Xie09} and which we found competitive (see Section \ref{sub:BgC}). 
The errors of estimation computed from the simulation data set $(S_2)$ are presented in SM, Section 7.4. Both methods are robust with respect to non-normal noise distribution,  and the normal-gamma BgC still offers a better quality of estimation than normexp-NP when the noise distribution departs from normality. \\

In conclusion, the normal-gamma background correction globally offers a better quality in signal estimation with respect to the normexp methods. Nevertheless, this improvement depends on the scale considered and does not steadily hold over the range of intensities.

\subsection{Operating characteristics}

Beyond the quality of estimation of the signal, the performance of a BgC procedure in practical experiments depends on its characteristics in terms of bias and variance.  In this section, we compare the operating characteristics of the normal-gamma and normexp BgC both on simulated and spike-in data. The results are gathered in Figure \ref{Fig4}. The background subtraction leading to probe deletion is not further considered. The data from $(E_3)$ are background corrected with the methods 1 to 4 described in Section \ref{BgC}. Quantile normalization based on both regular and negative probe intensities is applied, followed by log-transformation. 
The same procedures are implemented on the simulation data sets $(S_3)$ and $(S_4)$.

\subsubsection{Bias-precision trade-off}

The quality of a pre-processing method in microarray experiments can be characterised by its ability to distinguish between distinct values of the signal. Most of the procedures underestimate the signal fold-changes. This bias in fold-change estimation, called compression, has a negative impact on differential analysis. But the efficiency of a pre-processing method also depends on its precision, characterised by the variations of the corrected intensity for a given value of the signal. The trade-off between bias and precision is an indicator of the performance of a procedure. This issue can be understood by the example of a t-test statistic for a given probe differentially expressed between two groups: an important compression attenuates the difference of average intensities between the two groups, whereas a poor precision generates a high variance term in the denominator, which reduces the value of the test statistic.  \\

The compression and precision obtained with the four BgC methods  on the data set $(E_3)$ are presented on the first row of  Figure \ref{Fig4}. The first column displays the average intensity over the 34 spike bead types for each spike concentration. A similar saturation effect is observed for large concentrations with the four methods: for concentrations larger than 100pm, the relationship between  log-intensity and log-concentration is not linear.  Moreover, as the concentration decreases, a compression of the signal is observed with all the methods, but is significantly less pronounced with the normal-gamma BgC.

 The second column presents the average standard deviation between replicates over all spike bead types. We observe that the improvement in bias brought by the normal-gamma model is at the cost of a poorer precision. More generally,  the precision increases with the compression for the four methods. 

\subsubsection{Innate offset}

Shi et al \cite{Shi10} highlight the role played by the "innate offset", defined as the typical intensity assigned to the non-expressed genes by a pre-processing procedure, in the unbalanced bias-variance trade-off of the various BgC methods: the strategies which show the smallest innate offset usually generate less bias but present a poorer precision. On spike-in data sets, the innate offset is defined as the mean of the intensities measured on spike probes with concentration zero. The results displayed in Table \ref{Tab6} confirm the observations from Shi et al \cite{Shi10}: as underlined above, the normal-gamma BgC exhibits the smallest precision associated with the smallest bias with a slope from the linear regression of intensities on log-concentrations close to 1. The largest offset combined with the highest precision and the largest bias are observed for normexp-MLE. 

Shi et al \cite{Shi10} propose to compare the pre-processing methods in a more equal way by adding an offset to the background-corrected quantile-normalised intensities before log-transformation, in order to align the innate offsets of the various pre-processing strategies.  Our results  presented in SM, Section 7.5.1, indicate that the characteristics between the four BgC present more similarity after equalizing the innate offsets, but a slight difference remains between normexp-MLE and the other BgC methods on small intensities. Nevertheless, prior to the offset equalisation, the methods studied in this paper do not present the large range of bias-precision trade-offs observed in the pre-processing strategies considered in \cite{Shi10}. In this context, the equalisation of  the innate offsets does not appear sufficient to completely erase the differences between BgC methods.

\subsubsection{Operating characteristics on simulated data}

In order to reinforce the validation of the normal-gamma parametrisation for the noise-signal distribution, we compare the operating characteristics obtained on spike-in data to the ones provided by the normal-gamma simulated data from set $(S_3)$.  The spike concentration, used as references to assess the bias and precision of the procedures on spike-in data, is replaced by the true value of the signal.
The results are displayed on the second row of  Figure \ref{Fig4}.
The first column presents the average intensity as a function of the signal log-intensity, and the standard deviation of the replicates is shown in the second column. The trends are very similar to the ones observed on spike-in data, with a small difference for the variance with normexp-RMA.
Besides, we observe that the compression in small intensities generated by the four BgC methods is purely a statistical effect. However, the signal attenuation in high intensities observed on spike-in data is not present on simulated data. Indeed, it has already been suggested that this phenomenon could come from saturation in light intensity on microarrays.

Furthermore, we address the departure from normality observed on the negative probe distribution by simulating microarrays with a gamma distributed signal and a non-normal background noise (data set $(S_4)$). In order to get a realistic noise distribution, the background noise and the negative probe intensities are sampled from the quantile-normalised negative probe intensities from all arrays in $(E_3)$ (see details in SM, Section 7.1.2). The operating characteristics of the four BgC methods are presented on the third row of Figure \ref{Fig4}. We observe that the slight difference between normal-gamma simulations and spike-in data with normexp-RMA,  observed on the second row of Figure \ref{Fig4}, is partially corrected by generating a non-normal background noise.

The same quantities are computed based on normal-exponential simulated data with parameter sets 8 and 9. The results are displayed on the fourth row of Figure \ref{Fig4} for set 9, and on Figure E in SM for set 8. The comparison between the operating characteristics of the four BgC methods are absolutely not consistent with the observations from the spike-in data. In particular, 
the normal-exponential simulated data provide almost identical bias and precision curves for normal-gamma, normexp-NP and normexp-MLE, whereas these methods exhibit notable differences on spike-in data.  
\\

  The parallel drawn between the operating characteristics of the four BgC methods on spike-in and simulated data confirms that the gamma model represents a much more accurate parametrisation for the signal distribution than the usual exponential model.   \\

\subsection{Differential expression analysis}

The BgC methods are compared from a practical point of view through a differential expression analysis performed on the dilution data set $(E_4)$,  based on the hierarchical linear model approach from Smyth \cite{Smyth04} implemented in the {\tt limma} package. This procedure provides p-values from a moderated t-statistic. A first analysis is run on the two pure samples (proportions 100\%/0\% and 0\%/100\%) to define the "true" differentially expressed (DE) and non-differentially expressed probes. A second differential expression analysis performed on the two mixed samples (proportion 75\%/25\% and 25\%/75\%) is used to assess the performance of the BgC methods. The moderated t-test statistic implemented in the {\tt limma} package includes a variance term representing the variation of the gene intensity across all arrays, as well as hyperparameters computed from the whole data set intensities. Therefore, in order to get independent results, the two analyses are performed on separate linear models. 

The estimate proportion of DE probes in pure samples computed with a convex decreasing density procedure  \cite{Langaas05} is $28\%$ for all methods. Thus, in order to be conservative, we define the probes with the 20\% smallest p-values as "true DE", and the probes with the 40\% highest p-values as "true non-DE". 
Moderated t-statistic values are then computed from the comparison of the two mixed samples. The p-values from true DE and non-DE probes are ordered. The area under the ROC curve (AUC) is used to quantify the sensitivity of each BgC method, the largest value of the AUC corresponding to the highest sensitivity.  The four methods present similar AUC values but the normal-gamma BgC is slightly less competitive.

A similar analysis is run with the addition of an offset prior to log-transformation. Figure \ref{Fig5} displays the AUC values as a function of the added offset with each BgC method. The values observed for an offset equal to zero correspond to the sensitivity when a simple log-transformation is applied. As already highlighted by Shi et al \cite{Shi10}, we observe that the addition of a moderate offset increases the sensitivity of all BgC methods. For any value of the offset, the normexp methods outperform the normal-gamma.  The results obtained with the different normexp BgC methods are very similar, but we note that the highest sensitivity is achieved with normexp-RMA for offsets smaller than 50, and with normexp-MLE for offsets larger than 50.  \\

The BgC methods can also be compared regarding their ability to order a set of measured intensities corresponding to increasing or decreasing probe concentrations. This framework can refer, for example, to a longitudinal study where the gene expression is repeatedly measured at different times. The correlation between the mixture proportion and the intensity is analysed on the dilution data set $(E_5)$. For the true DE probes, the intensity is expected to be increasing or decreasing with the proportion. 

The dilution data sets  $(E_4)$ and  $(E_5)$ are based on the same pure biological samples. Therefore, true DE and non-DE probes defined on $(E_4)$ can be considered in the analysis of the data from $(E_5)$. The BeadChips used in experiments $(E_4)$ and $(E_5)$ are different, but some bead types are present on both devices. By mapping the annotation files from both BeadChips, the sets of probes respectively defined as DE and non-DE on $(E_4)$, and present on $(E_5)$ are extracted.

For each probe, the Spearman correlation coefficient is computed between the vector of mixture proportions and the observed intensities. This provides a test statistitic based on the ranking of the background corrected intensities, which allows a comparison of the BgC methods independently from the scale at which the data are analysed, provided that the transformation applied to the data is increasing.
 In particular, the results are not affected by the addition of an offset. The correlation coefficient is computed  separately on microarrays with starting RNA quantities 250ng, 100ng, 50ng  and 10ng. The coefficient is expected to be close to 1 in absolute values for the DE probes, and close to 0 for the non-DE. The probes are ranked according to their correlation coefficient value, and the resulting AUCs for each starting RNA quantity are displayed in Table \ref{Tab7}. We observe that the normal-gamma BgC is slightly but steadily less sensitive that the other methods.
 The AUCs observed with the different normexp estimates are very similar and do not allow to assess the superiority of one method over the others.

\section{Discussion}\label{sec:discussion}

In many microarray experiments, background noise correction is an important issue in order to improve the measurement precision. Model-based background correction procedures have been developed as an alternative to the default background subtraction from Illumina BeadStudio which has proved to remove informative probes. The usual normal-exponential model considered for the noise-signal distribution has already been pointed out as inappropriate for Illumina BeadArrays \cite{Wang11}. Our observations confirm this result by highlighting the poor fitting of the normexp reconstructed densities on observed intensities with three different parameter estimates . 
We propose an alternative model based on a more flexible parametrisation of the signal which is assumed to follow a gamma distribution, as well as the associated background correction. The reconstructed density offers a better fit of the distribution of the observed intensities, validating this new model as more appropriate for Illumina microarrays.  \\

We compare the performance of the background correction procedures based on the normal-gamma and normal-exponential models, on simulated and experimental data sets. 
Our simulation study indicates that the normal-gamma model brings an overall improvement in terms of signal estimation, characterised by a smaller average difference between the true signal and the background corrected intensity. But surprisingly, the differential expression analysis run on two dilution data sets shows that the improvement in terms of parametrisation does not have a positive impact on practical experiments. 
This result may be explained in two ways. 

On one side, the operating characteristics of the background correction procedures  are compared on a set of spike-in data, which allow to connect the probe intensity with the concentration of the target gene in the biological sample. We note that the normal-gamma model generates less bias than the normexp methods, but at the cost of a loss in precision. With the addition of an offset prior to the log-transformation, which provides balance in the bias-precision trade-off of the different methods, the operating characteristics appear similar, suggesting comparable performance.

On the other side, we examine the error in signal estimation as a function of the signal on log-scale simulated data. The normal-gamma model outperforms the other methods on small intensities, but is less competitive on moderate intensities. 
Due to the marked compression of the recovered intensity when the signal decreases, the improvement in terms of signal estimation for the small intensities has a weak effect on the differential expression analysis. Thus, the smaller average error of estimation observed with the normal-gamma background correction does not result  in a higher sensitivity in practical experiments. \\

Besides, the parallel drawn between the operating characteristics of the different background corrections obtained, on the one hand with spike-in data and on the other hand with normal-gamma simulated data, highlights high similarities. The simulations from the normal-gamma model recover subtile differences between background correction procedures, whereas simulations from the normexp model totally fail to reproduce the trends noticed on spike-in data. These considerations
enhance the validation of the normal-gamma model for Illumina microarrays, and illustrate the potential of the normal-gamma simulations for the comparison of pre-processing procedures. Furthermore, the similarities between the observations from spike-in and simulated data are increased by sampling the background noise from the empirical negative probe distribution which suggests that an improvement in modeling  could be brought by a non-normal parametrisation of the background noise. \\

In conclusion, this paper addresses the lack of fit of the usual normal-exponential model by proposing a more flexible parametrisation of the signal distribution as well as the associated background correction. 
This new model proves to be considerably more accurate for Illumina microarrays, but our results indicate that the improvement in terms of modeling does not lead to a higher sensitivity in differential analysis. Nevertheless, this realistic modeling makes way for future investigations, in particular to examine the characteristics of pre-processing strategies.

\section{Figures and Tables}

\medskip

 \begin{figure}[!h]
  \begin{center}
  \includegraphics[width=5cm,height=5cm, trim= 35mm 35mm 35mm 35mm]{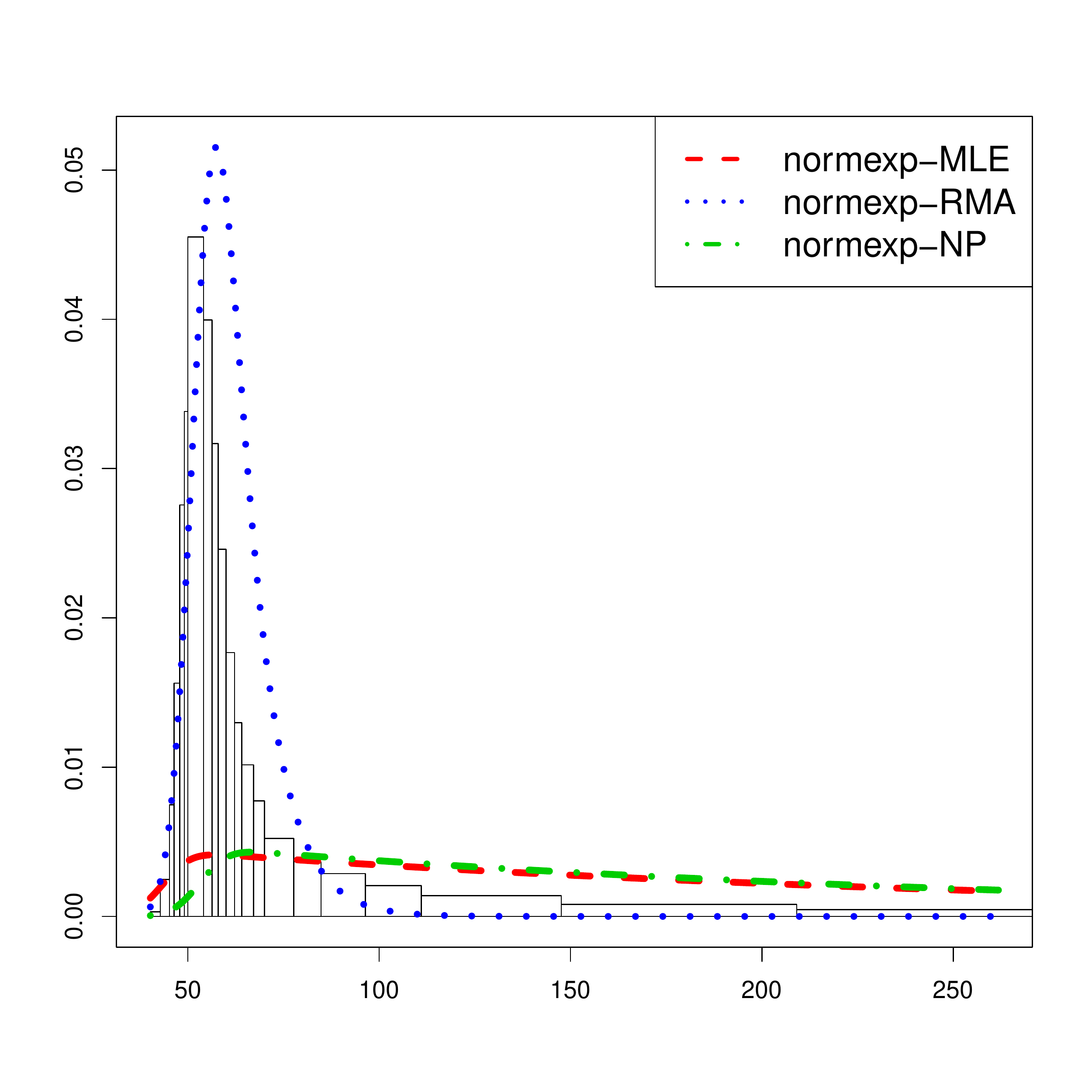}
   \end{center}
  \caption{ Normal-Exponential estimation for one array from $(E_1)$ after removal of imperfectly designed probes: irregular density histogram of all regular probe intensities and the {\it plug-in} normexp density of the regular probes with MLE,  RMA  and NP parameter estimates.}
   \end{figure}
  \label{Fig1}

\begin{figure}[!h]
\begin{center}
  \includegraphics[width=5cm,height=5cm, trim= 35mm 35mm 35mm 35mm]{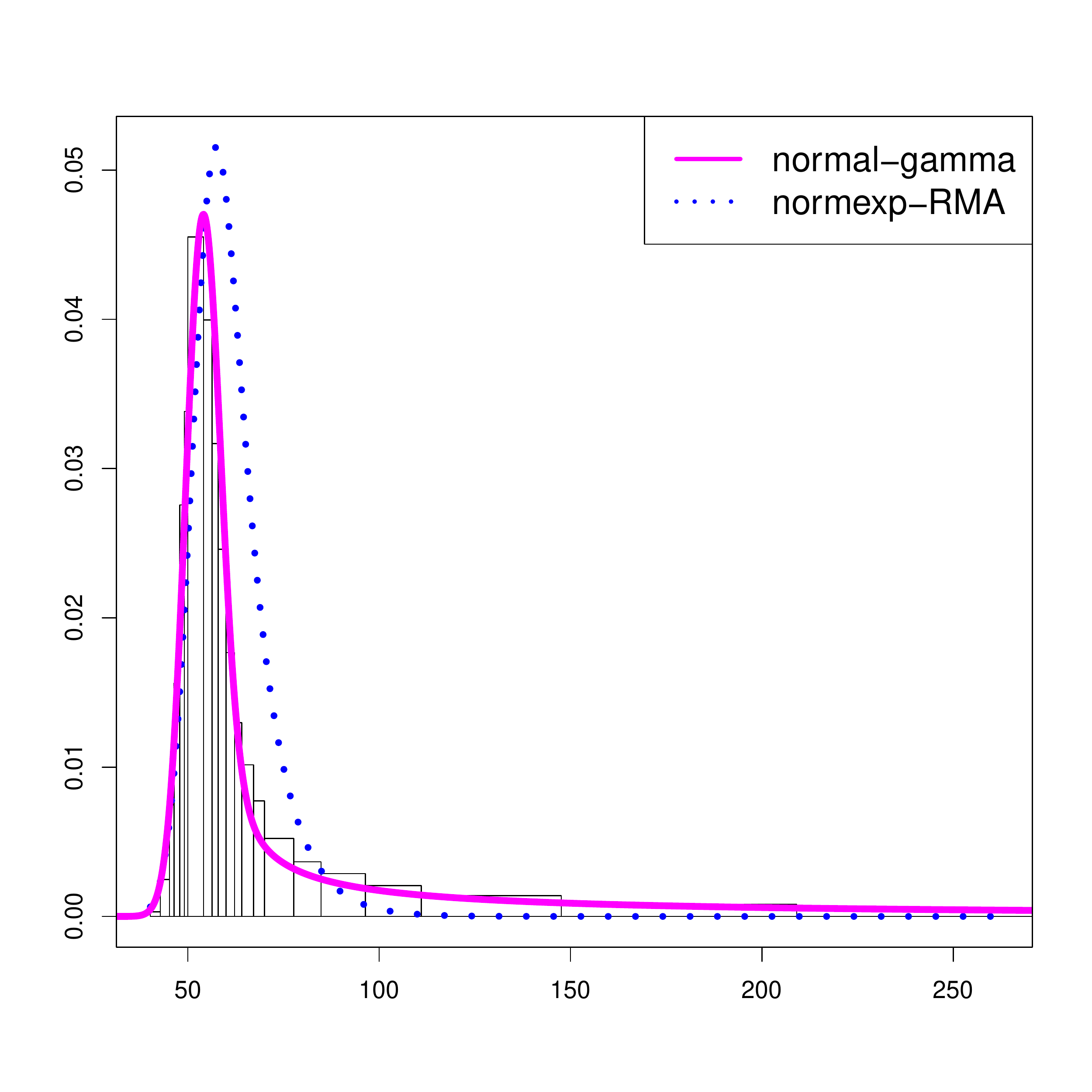}
  \end{center}
  \caption{Normal-Gamma estimation for one array from $(E_1)$ after removal of imperfectly designed probes: irregular density histogram of all regular probe intensities, {\it plug-in} normexp density with RMA estimate
and {\it plug-in} normal-gamma density with MLE estimate.}
\label{Fig2}
  \end{figure}

\begin{figure}
\begin{center}
 \includegraphics[width=9cm,height=15cm]{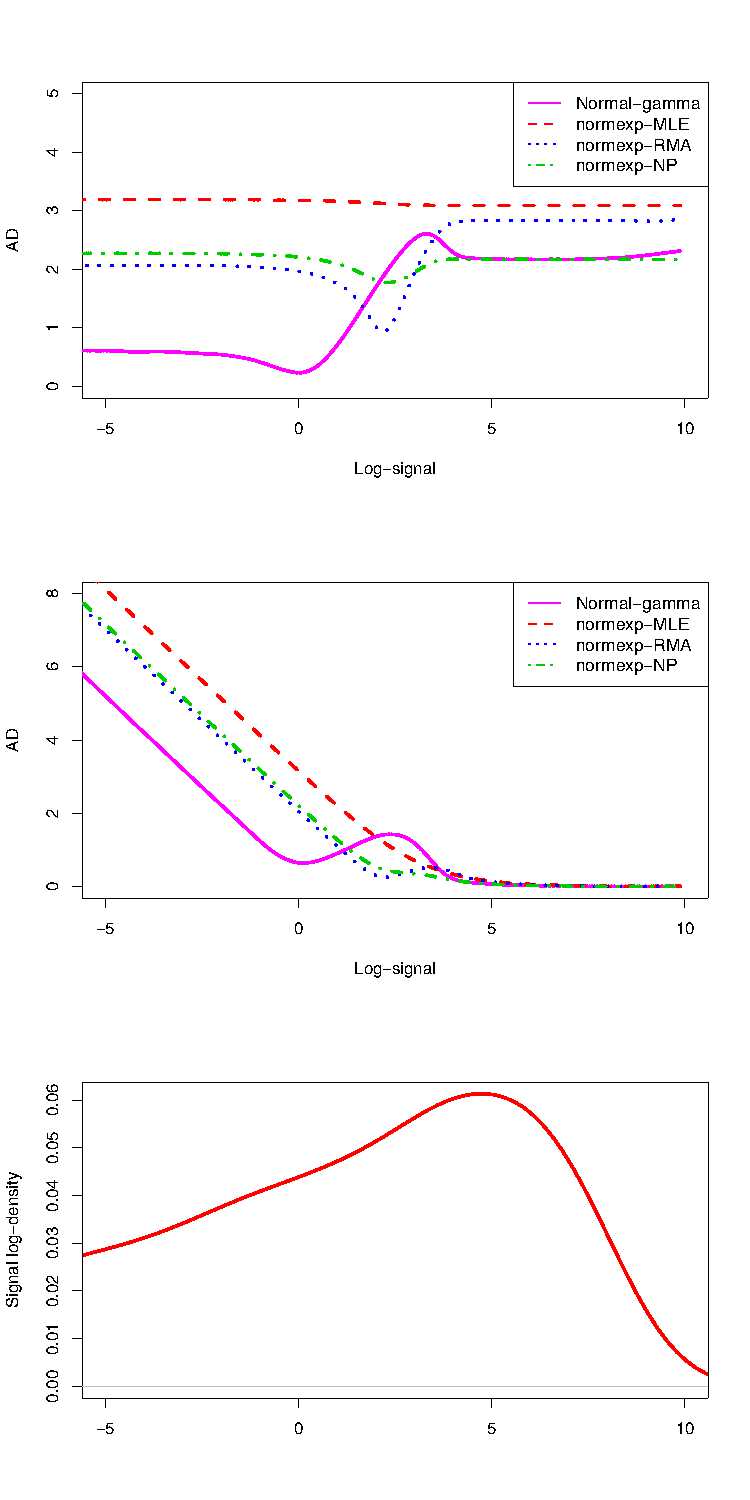}
  \end{center}
  \caption{ Logarithm of the Absolute Deviation of estimated signal on raw scale (first row), Absolute Deviation of log-transformed estimated signal (second row) and signal log-density (third row). Normal-gamma BgC (purple) and normexp BgC with MLE (pink), RMA (blue) and NP (green) parameters.}
  \label{Fig3}
  \end{figure}

\begin{figure}
\begin{center}
  \includegraphics[width=10cm,height=20cm]{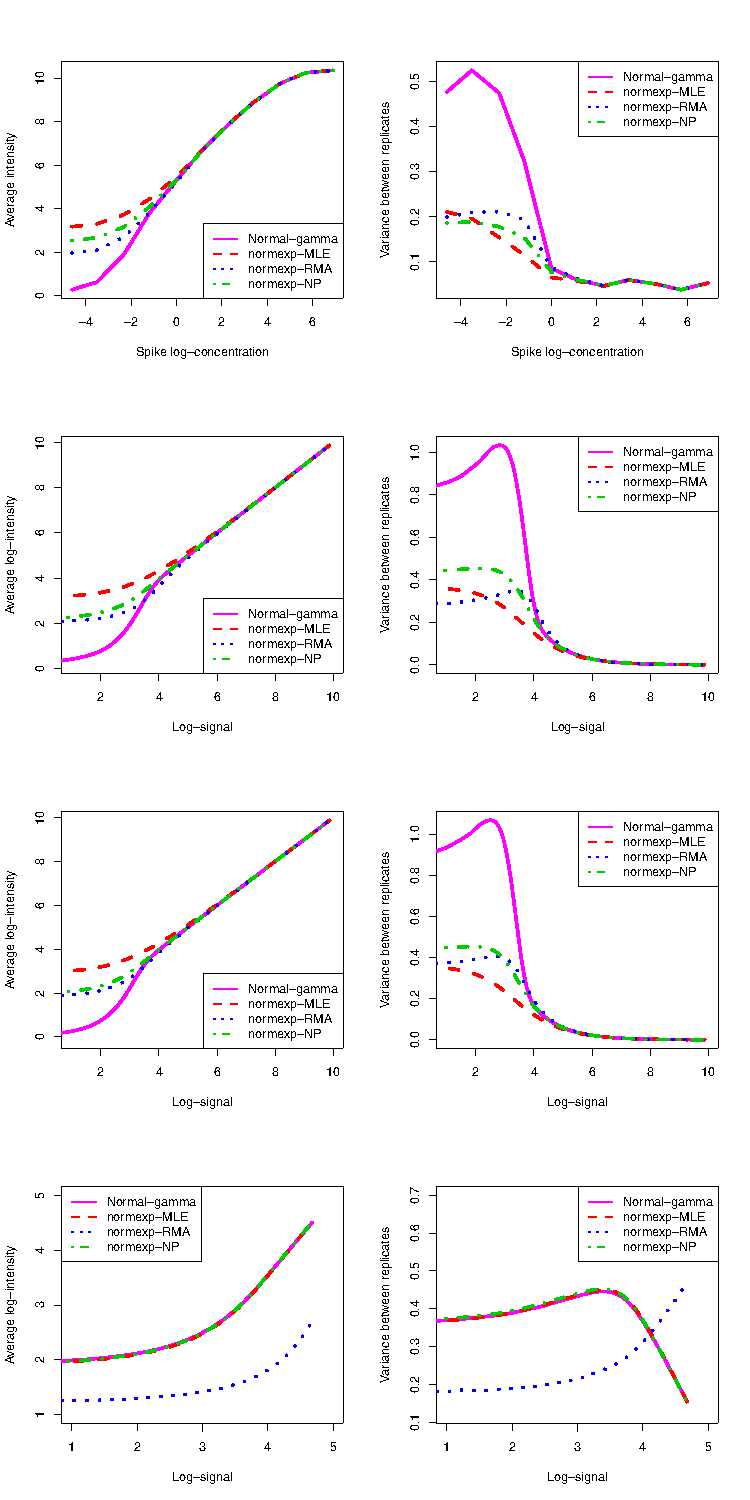}
  \end{center}
  \caption{ Operating characteristics of the BgC methods on spike-in and simulated data.  Row 1: average spike intensities (left) and standard deviation of spike replicates (right) for all non-zero spike concentrations. Row 2 to 4: average intensity (left) and standard deviation of replicates (right) as a function of signal intensity. Row 2: normal-gamma simulation in data set $(S_3)$  (parameter set 7); Row 3: gamma signal and empirical background noise distribution (data set $(S_4)$); Row 4 normal-exponential simulation in data set $(S_3)$ (parameter set 9). }
 \label{Fig4} 
\end{figure}

\begin{figure}
\begin{center}
 \includegraphics[width=9cm,height=9cm]{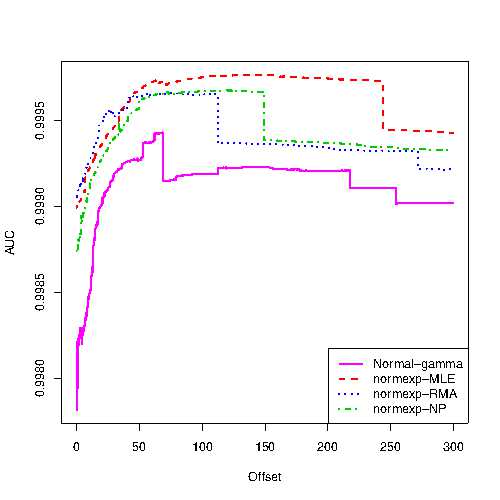}
  \end{center}
  \caption{ AUC as a function of added offset. AUC from moderated t-test for mixed sample differential analysis  in data set $(E_4)$ (proportion 25\%/75\% and 75\%/25\%) for different values of offset.}
 \label{Fig5} 
\end{figure}

\newpage

 \begin{table} [!h]   
\begin{center}
    \mbox{
 \begin{tabular}{lccc}\hline
\rule{0pt}{15pt} \vspace{0.1cm} & Human  & Human & Mice  \\
 \rule{0pt}{15pt} \vspace{0.2cm} & (all probes) & (remove bad probes) & \\\hline
\rule{0pt}{15pt} \vspace{0.1cm} nexp MLE & 7.09& 5.14 & 4.83 \\
\rule{0pt}{15pt} \vspace{0.1cm} nexp RMA & 2.96 &  3.18 & 2.71 \\
\rule{0pt}{15pt} \vspace{0.1cm} nexp NP & 7.69 & 5.50& 5.29\\\hline
\rule{0pt}{15pt} \vspace{0.1cm} Abs Dev normgam & 0.17  & 0.21 & 0.20 \\\hline
\end{tabular}
	}
\end{center}
\caption{Average deviation between normexp reconstructed density and histogram divided by the deviation between normal-gamma reconstructed density and histogram (First row: RMA estimator; second row: MLE normexp estimator; third row: NP normexp estimator). The fourth row gives the mean deviation of the normal-gamma estimator as a reference. The mean is computed over the ten arrays from $(E_1)$ with (first column) or without (second column) the non specific binding probes and over the four arrays from $(E_2)$ (third column).}
\label{Tab1}
\end{table}

 \begin{table}[!h]     
 \begin{center}   
    \mbox{

\begin{tabular}{lcccc}\hline
 \rule{0pt}{15pt} \vspace{0.2cm} & $\mu$ & $\sigma $ & $k$ & $ \theta$  \\\hline
\rule{0pt}{15pt} \vspace{0.1cm}  set 1 & 7.1\sc{e}-4 & 5.6\sc{e}-3 & 9.3\sc{e}-3 & 1.7\sc{e}-2 \\
 \rule{0pt}{15pt} \vspace{0.1cm}   set 2 & 1.3\sc{e}-3 & 5.5\sc{e}-3 & 1.0\sc{e}-2 & 1.8\sc{e}-2 \\
 \rule{0pt}{15pt} \vspace{0.1cm}   set 3 & 3.5\sc{e}-3 & 1.6\sc{e}-2 & 6.9\sc{e}-3 & 8.3\sc{e}-3 \\
 \rule{0pt}{15pt} \vspace{0.1cm}   set 4 & 4.5\sc{e}-3 & 1.3\sc{e}-2 & 8.9\sc{e}-3 & 9.8\sc{e}-3   \\
\rule{0pt}{15pt} \vspace{0.1cm}    set 5 & 2.1\sc{e}-3& 7.6\sc{e}-3 & 2.6\sc{e}-2 & 1.7\sc{e}-2 \\
\rule{0pt}{15pt} \vspace{0.1cm}    set 6 & 3.5\sc{e}-3 & 7.2\sc{e}-3 & 3.9\sc{e}-2 & 2.4\sc{e}-2 \\\hline
   \end{tabular}

	}
\end{center}
\caption{ Relative $L_1$-error for each parameter in the normal-gamma model using MLE estimates.}
\label{Tab2}
\end{table}

\begin{table}[!h]
 \begin{center}   
    \mbox{
\begin{tabular}{lccc}\hline
\rule{0pt}{15pt} \vspace{0.2cm} & $\mu$ & $\sigma $ & $ \theta$  \\\hline
 \rule{0pt}{15pt} \vspace{0.1cm}   set 3 & 1.1 & 1.0 & 1.6 \\
 \rule{0pt}{15pt} \vspace{0.1cm}  set 4 & 1.2 & 1.0 & 1.8   \\
 \rule{0pt}{15pt} \vspace{0.1cm}  set 5 & 2.1 &1.0  & 2.3 \\
 \rule{0pt}{15pt} \vspace{0.1cm}  set 6 &1.9 & 1.0 &  1.9\\\hline
   \end{tabular}

}
\end{center}
\caption{ Error in parameter estimation. Ratio between the relative $L_1$ errors of the MLE estimation in the normal-gamma and in the normexp models for $(\mu,\sigma,\theta)$ from normexp data.}
\label{Tab3}
\end{table}

\begin{table}[!h]
 \begin{center}   
    \mbox{
    
\begin{tabular}{ccccccc}\hline
\rule{0pt}{15pt} \vspace{0.2cm}$(\mu , \sigma,k ,\theta )$&  $R(1)$ & $R(2)$ & $R(3)$ &  $R(4)$ & $R(5)$ & $\mathrm{MAD}(\widehat S^{(0)})$ \\
\hline
\rule{0pt}{15pt} \vspace{0.1cm}set 1 &  1.00 & 4.16 & 1.77 & 1.52  & 1.16 & 2.34   \\ 
\rule{0pt}{15pt} \vspace{0.1cm}set 2 &  1.00 & 4.10 & 1.90 & 1.66 & 1.20 & 11.7 \\
\rule{0pt}{15pt} \vspace{0.1cm}set 3 &  1.00 & 1.00 & 4.69 & 1.00 & 1.00 &  4.57\\
\rule{0pt}{15pt} \vspace{0.1cm}set 4 &  1.00 & 1.00 & 3.71 & 1.00 & 1.02 & 31.4 \\
\rule{0pt}{15pt} \vspace{0.1cm}set 5 &  1.00 & 1.00 & 2.11 & 1.00 & 1.15 & 2.95\\
\rule{0pt}{15pt} \vspace{0.1cm}set 6 &  1.00 & 1.00 & 1.46 & 1.00 & 1.35 & 17.2 \\
\hline
\end{tabular}
}
\end{center}
\caption{Excess risk ratio of background corrected raw-scale intensities. Mean Absolute Deviation (MAD) of the background corrected intensities for methods $\widehat{S}^{(j)}$, $j=1,\dots,5$ divided by the MAD for the theoretical normal-gamma BgC with the true parameters (method  $\widehat{S}^{(0)}$), from the simulation data set $(S_1)$. Column 1: normal-gamma, column 2: normexp-MLE, column 3: normexp-RMA, column 4: normexp-NP, column 5: background subtraction.
 The MAD of the theoretical normal-gamma deconvolution with the true parameters is given as reference in column 6. }
\label{Tab4}
\end{table}

\begin{table}[!h] 
 \begin{center}   
    \mbox{
    
\begin{tabular}{ccccc}\hline
\rule{0pt}{15pt} \vspace{0.2cm}$(\mu , \sigma,k ,\theta )$&  $\text{R(1)}$ & $\text{R(2)}$ & $\text{R(3)}$ &  $\text{R(4)}$  \\\hline
\rule{0pt}{15pt} \vspace{0.1cm}set 1 &  1.00 & 1.32 & 1.18 & 1.17    \\ 
\rule{0pt}{15pt} \vspace{0.1cm}set 2 &  1.00 & 1.28 & 1.16 & 1.16  \\
\rule{0pt}{15pt} \vspace{0.1cm}set 3 &  1.00 & 1.00 & 2.98 & 1.00 \\
\rule{0pt}{15pt} \vspace{0.1cm}set 4 &  1.00 & 1.00 & 2.45 & 1.00 \\
\rule{0pt}{15pt} \vspace{0.1cm}set 5 &  1.00 & 1.00 & 1.81 & 1.00 \\
\rule{0pt}{15pt} \vspace{0.1cm}set 6 &  1.00 & 1.00 & 1.39 & 1.00 \\\hline
\end{tabular}

}
\end{center}
\caption{Excess risk ratio of background corrected log-transformed intensities. Mean Absolute Deviation (MAD) of the background corrected intensities for methods $\widehat{S}^{(j)}$, $j=1,\dots,5$ divided by the MAD for the theoretical normal-gamma BgC with the true parameters (method  $\widehat{S}^{(0)}$), from the simulation data set $(S_1)$. Column 1: normal-gamma, column 2: normexp-MLE, column 3: normexp-RMA, column 4: normexp-NP. }
\label{Tab5}
\end{table}

\begin{table}[!h]
 \begin{center}   
    \mbox{

 \begin{tabular}{cccc}\hline
\rule{0pt}{15pt} \vspace{0.2cm}BgC & Innate offset  &  Stand. Dev. & Slope  \\\hline
\rule{0pt}{15pt} \vspace{0.1cm}normexp MLE &23.4 & 0.095 & 0.74 \\
\rule{0pt}{15pt} \vspace{0.1cm}normexp NP & 12.4 & 0.100& 0.80 \\
\rule{0pt}{15pt} \vspace{0.1cm}normexp RMA& 6.9  & 0.110 & 0.86 \\
\rule{0pt}{15pt} \vspace{0.1cm}normal-gamma& 1.5 & 0.200 & 0.99 \\\hline
\end{tabular} 
}
\end{center}
\caption{ Innate offset, average standard deviation of spike replicates and slope of the linear regression of the spike average intensity on the log-concentration. }
\label{Tab6}
\end{table}

\begin{table}[!h]
 \begin{center}   
    \mbox{
 \begin{tabular}{ccccc}\hline
\rule{0pt}{15pt} \vspace{0.1cm} & Normal-  & Normexp- & Normexp- & Normexp- \\
\rule{0pt}{15pt} \vspace{0.2cm} & gamma & MLE & RMA & NP\\\hline
\rule{0pt}{15pt} \vspace{0.1cm}250ng &0.9778& 0.9812& 0.9813 & 0.9820\\
\rule{0pt}{15pt} \vspace{0.1cm}100ng  & 0.9774 & 0.9807& 0.9809& 0.9808 \\
 \rule{0pt}{15pt} \vspace{0.1cm}50ng&  0.9805  & 0.9834 & 0.9832& 0.9841 \\
\rule{0pt}{15pt} \vspace{0.1cm}10ng& 0.9782 & 0.9818 & 0.9787& 0.9816 \\\hline
\end{tabular}
}
\end{center}
\caption{AUC from Spearman correlation test between the proportion and the intensity in the dilution data set $(E_5)$, for the four BgC methods, and the four RNA starting concentrations.  }
\label{Tab7}
\end{table}

\newpage

\newpage

\section{Appendix: supplementary material}

\subsection{Experimental and simulated data sets} 

\subsubsection{Laboratory methods for NOWAC data set $(E_1)$}

Data set $(E_1)$  consists of the gene
expression profiles in peripheral blood of ten controls from the Norwegian Women And Cancer cohort.
The gene expression profiles were measured on source cells
using Illumina Human HT-6 v4 Expression BeadChip (Illumina, San Diego,
CA), which enables genome-wide expression analysis (more than 48 000 transcripts) of six samples in parallel on a single microarray. A restricted set of more than 25,000 probes is also considered by removing non-reliable probes according to Illumina's annotation files.  The Illumina TotalPrep RNA amplification Kit (Ambion Inc.,
Austin, TX) was used to amplify RNA for hybridization on Illumina BeadChips. To
synthesize the first strand cDNA by reverse transcription, we used
totalRNA from each sample collected above. Following the second strand
cDNA synthesis and cDNA purification step, the in vitro transcription
to synthesize cRNA was prepared overnight within 12 hours.
The microarray service was provided by NMC-NTNU, a Norwegian national technology platform supported by the functional genomics program (FUGE) of the Research Council of Norway.

\subsubsection{Simulation of microarray data}

\textbf{Simulation experiment $(S_1)$.}
Microarray data are simulated from both normal-gamma and normexp models. In order to get realistic values of the parameters, six sets of parameters are computed by applying normal-gamma MLE, as well as normexp MLE and RMA estimation methods to observed intensities from one microarray from $(E_1)$ and $(E_2)$. The values of the six sets of parameters are summarized in Table A (sets 1-6). 
For each set of parameters,  $N=100$  random arrays are generated under the normal-gamma model, with $n_{\text{reg}}=25000$ regular probes and $n_{\text{neg}}=1000$ negative probes. For each repetition $\ell = 1, \dots , N$, two independent samples $ \textbf{X}^{\ell}$ and $\textbf{X}^{0,\ell}$, corresponding respectively to the expression levels of the regular and negative probes, are generated:

 \begin{itemize}
  \item $ \textbf{X}^{\ell} = \{ X_j^\ell = S_j ^{\ell}+ B_j^{\ell}, j=1, \dots ,n_{\text{reg}}  \} $ with 
   \begin{itemize}
 \item[] $\{ S_j^{\ell}, j=1, \dots ,n_{\text{reg}}  \} $ independent identically distributed (i.i.d.) sample from a gamma distribution with shape $k$ and scale $\theta$.
 \item[] $\{ B_j^{\ell}, j=1, \dots ,n_{\text{reg}}  \} $ i.i.d. sample from a normal distribution with mean $\mu$ and variance $\sigma ^2$.
\end{itemize}
 \item $\textbf{X}^{0,\ell} =  \{ B_j^{0,\ell}, j=1, \dots ,n_{\text{neg}}  \} $ i.i.d. sample from normal distribution with mean $\mu$ and variance $\sigma ^2$.
\end{itemize} 

\textbf{Simulation experiment $(S_2)$.}
The procedure is similar to $(S_1)$, but the background noise and the negative probe intensities are generated from a mixture distribution

$$(1-p)\mathcal N(\mu  \sigma )+p\, \upchi^2(3,55)$$ 

\noindent where $\upchi^2(3,55)$ is a $\upchi^2$-distribution with 3 degrees of freedom and a non-centrality parameter equal to 55, for $p$ in $(0.1, 0.25, 0.5, 0.75, 1)$. The values of the normal-gamma parameters are given in Table A,  set 1. \\

\textbf{Simulation experiment $(S_3)$.}
The parameter values for this experiment, computed from one array in experimental data set $(E_3)$ are given in Table A: set 7 is computed from the normal-gamma model, and sets 8 and 9 are estimated from the normexp model with MLE and RMA parameter estimates.
In order to mimic replicates from the same biological sample, the vector of signal intensities $\textbf{S}$, drawn from a gamma distribution is identical on the $N$ arrays.
Then, for each repetition $\ell = 1, \dots , N$, two independent samples $ \textbf{B}^{\ell}$ and $\textbf{X}^{0,\ell}$ corresponding respectively to the intensities of the background noise on regular probes  and the intensities of negative probes are generated.  \\

 \begin{itemize}
  \item $ \textbf{X}^{\ell} = \{ X_j^\ell = S_j + B_j^{\ell}, j=1, \dots ,n_{\text{reg}}  \} $ with 
   \begin{itemize}
 \item[] $\{ S_j, j=1, \dots ,n_{\text{reg}}  \} $ = \textbf{S}.
 \item[] $\{ B_j^{\ell}, j=1, \dots ,n_{\text{reg}}  \} $ i.i.d. sample from a normal distribution with mean $\mu$ and variance $\sigma ^2$.
\end{itemize}
 \item $\textbf{X}^{0,\ell} =  \{ B_j^{0,\ell}, j=1, \dots ,n_{\text{neg}}  \} $ i.i.d. sample from a normal distribution with mean $\mu$ and variance $\sigma ^2$.
\end{itemize}

Finally, $\textbf{X}^{\ell}$ and $\textbf{X}^{0,\ell}$ stand for the regular and negative probe intensities on array $\ell$.  Note that the parameter sets 8 and 9 with a shape value of 1 correspond to an exponential distribution. \\

\textbf{Simulation experiment $(S_4)$.}
This data set is simulated based on a gamma distributed signal and a non-normal background noise. In order to get a realistic distribution of the background noise, the negative probe intensities from the 48 arrays in experimental data-set $(E_3)$ are quantile-normalized independently from the regular probes, and gathered in a vector $\textbf{D}_{\text{neg}}$. The signal identical over all arrays is simulated from a gamma distribution with parameters $k$ and $\theta$ from set 7. Then, for each array $\ell$, two vectors $ B^{\ell}$ and $\textbf{X}^{0,\ell}$ of length $n_{\text{reg}}$ and $n_{\text{neg}}$ standing for the background noise and the negative probe intensities  are sampled  with replacement from $\textbf{D}_{\text{neg}}$.

\begin{table}
\begin{center}
\begin{tabular}{ccccccr}\hline 
\rule{0pt}{15pt} \vspace{0.2cm}& $\mu$ & $\sigma$ & $k$ & $\theta$ &Exp. data set  & est. method  \\\hline
\rule{0pt}{15pt} \vspace{0.1cm} set 1 & 53 & 4.4 & 0.12 & 1785 & $(E_1)$ & $f^{\text{ng}}$ + MLE\\
\rule{0pt}{15pt} \vspace{0.2cm} set 2 & 138 & 24 & 0.11 & 4949 & $(E_2)$&$f^{\text{ng}}$ + MLE\\ 
\rule{0pt}{15pt} \vspace{0.1cm} set 3 & 43.5 & 5.8 & 1 & 226 & $(E_1)$ & $f^{\text{nexp}}$ + MLE \\
\rule{0pt}{15pt} \vspace{0.2cm} set 4 & 170 & 41 & 1 & 505 & $(E_2)$ &  $f^{\text{nexp}}$ + MLE\\
\rule{0pt}{15pt} \vspace{0.1cm} set 5 & 52.8 & 5.0 & 1 & 8.33 & $(E_1)$ &$f^{\text{nexp}}$ + RMA\\
\rule{0pt}{15pt} \vspace{0.2cm} set 6 & 223 & 37 & 1 & 33.8 & $(E_2)$ & $f^{\text{nexp}}$ + RMA \\
\rule{0pt}{15pt} \vspace{0.2cm} set 7 & 93 & 11 & 0.08 & 3230 & $(E_3)$ & $f^{\text{ng}}$ + MLE\\
\rule{0pt}{15pt} \vspace{0.2cm} set 8 &69  & 13 & 1 & 277 & $(E_3)$ & $f^{\text{nexp}}$ + MLE\\
\rule{0pt}{15pt} \vspace{0.2cm} set 9 & 92 & 14 & 1 & 10.5 & $(E_3)$& $f^{\text{nexp}}$ + RMA\\\hline
\end{tabular}
\end{center}
\caption*{Table A. The nine sets  of parameters used in the simulations by using on array $(E_1)$, $(E_2)$ and $(E_3)$ with the three methods of estimation. The $k=1$ value corresponds to a normexp model with an exponential distribution with mean $\theta$.}
 \end{table}

\newpage

\subsection{Fit of normexp and normal-gamma model on experimental data sets $(E_1)$ and $(E_2)$}

\subsubsection{Fit of normal model on negative probe distribution}

On several arrays from $(E_1)$ and $(E_2)$, we have compared the density histogram of negative probes to the {\it plug-in} normal density $f^{\text{norm}}_{\hat\mu,\hat\sigma}$ obtained by using robust estimators of the parameters 
$(\mu, \sigma)$:

\begin{equation*}
\left\{ \begin{array}{l}
\widehat{\mu} = \text{median} ( \{  X_j , j\in J_0 \} )  \\
\widehat{\sigma} = \text{median}( \{  \left|  X_j - \widehat{\mu}\right| , j\in J_0 \} )/ 0.6745.
\end{array} \right. 
\end{equation*}
 
The results of this comparison, similar through the arrays, are illustrated in Figure B on one array from $(E_1)$ and $(E_2)$. 
This figure presents a regular density histogram of the negative probe intensities as well as the {\it plug-in} normal density. The empirical distribution is essentially normal but we notice a slightly heavier right tail. 
\vspace{1cm}

\begin{figure*}[!h]
\begin{center}
 \includegraphics[height=7cm, width=14cm]{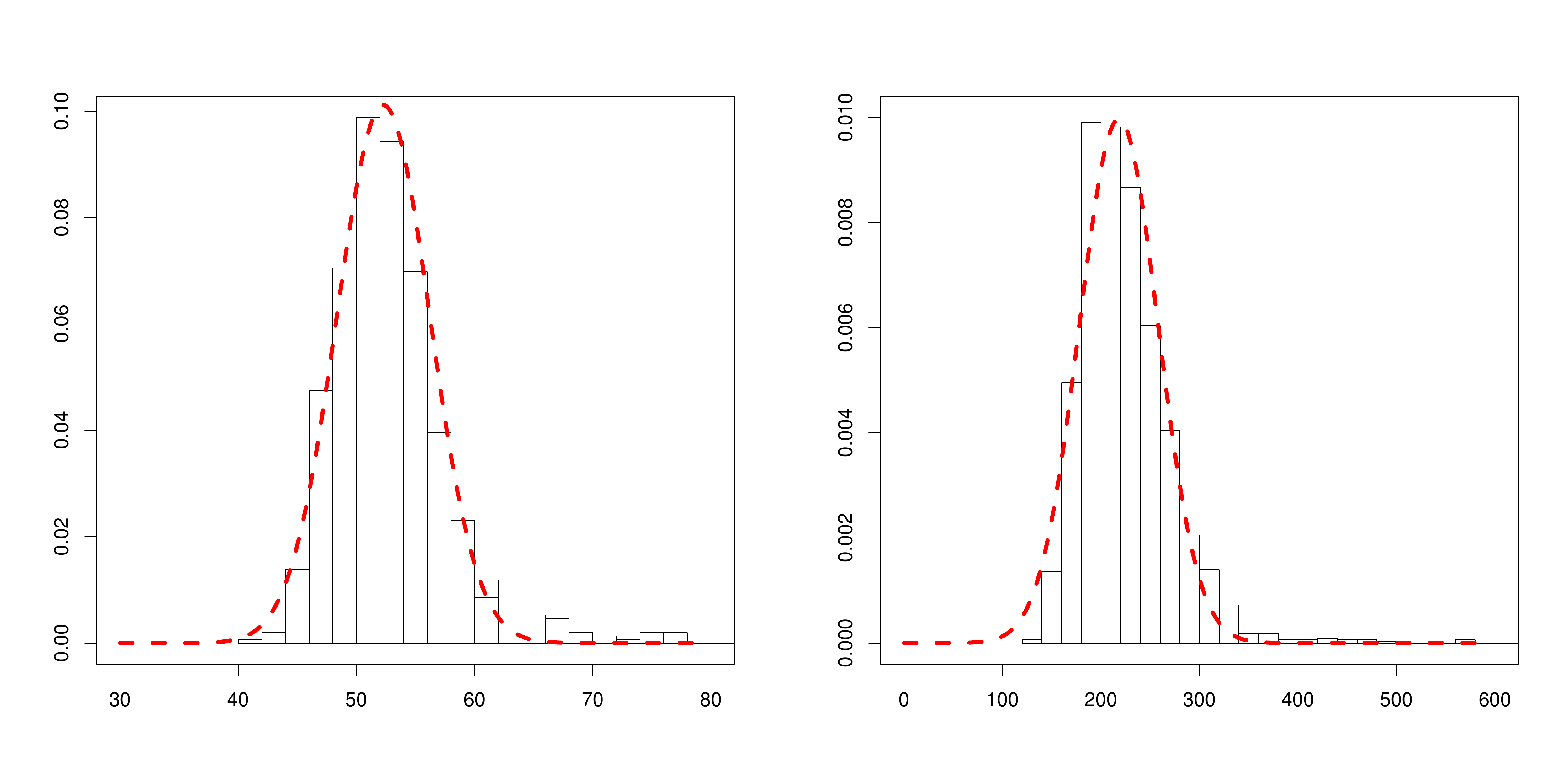} 
\caption*{Figure A.  Regular density histogram of the negative probe intensities and {\it plug-in} normal density using a robust estimate of the parameters (red dotted line) built on the negative probes. Left: one array from $(E_1)$; right: one array from $(E_2)$. \vspace{0.5cm}}
\end{center}
\end{figure*}

\subsection{Fit of the normexp and normal-gamma models.}

Figure B presents the fit of the normal-gamma and normexp models with various parameter estimates on two arrays from $(E_1)$ and $(E_2)$.

\begin{figure}
\begin{center}
\includegraphics[height=15cm, width=15cm]{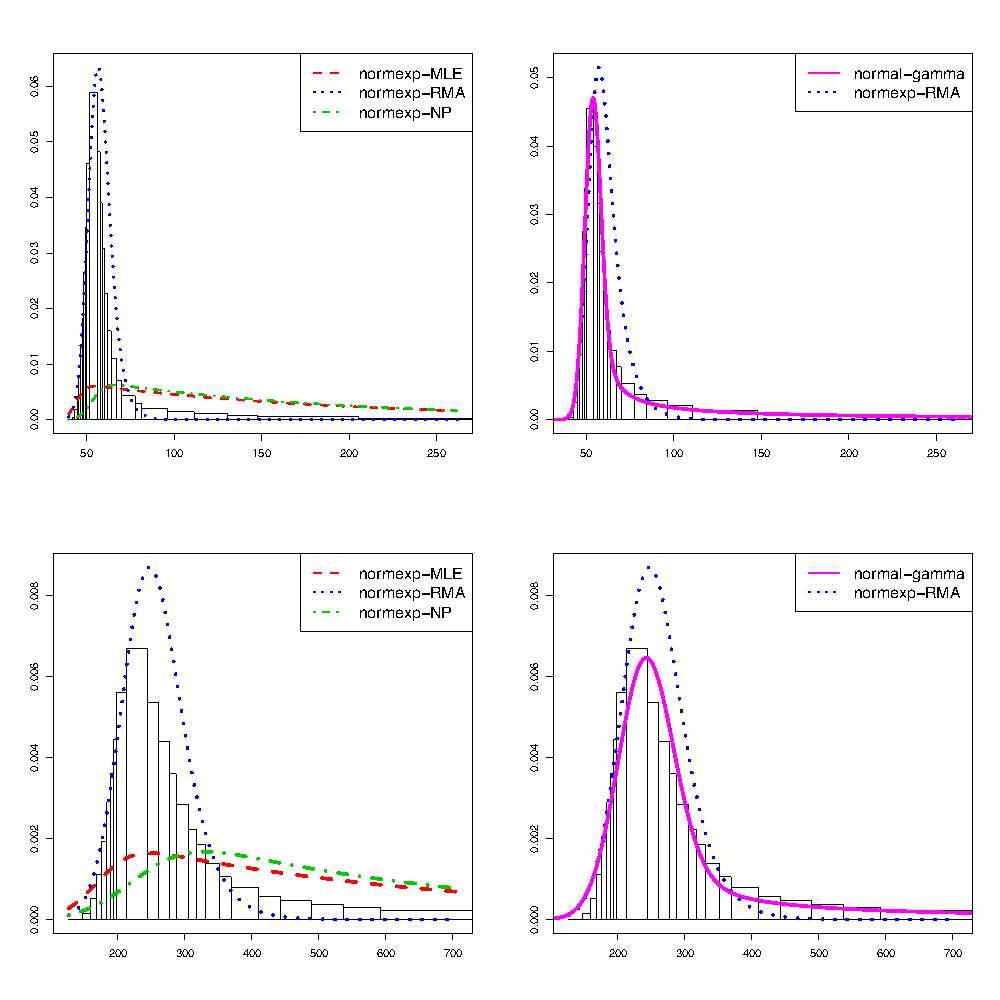}
\end{center}
\caption*{Figure B. Normal-gamma and normal-exponential  fit on one array from $(E_1)$ with the full set of regular probes (first row) and one array from $(E_2)$ (second row). On each figure the irregular density histogram of the regular probe intensities is displayed. On left column are presented the plug-in normexp density with MLE, RMA and NP parameter estimates, and on the right column the plug-in normal-gamma density estimator, together with the normexp-RMA estimator. }
\end{figure}

\newpage

\subsection{Quality of the signal estimation at the log-scale on the data set $(S_3)$ (set 7)} 

For each background correction procedure $\widehat{S}^{(i)}$ $i=0,\ldots,4$ described in Section 6.2.4, we compute the Mean Absolute Deviation of the bakground corrected log-transformed intensities over $N=100$ repetitions. Note that the background corrected values after background subtraction include negative values and therefore can not be log-transformed.
For each parameter set 1 to 6 and for each BgC methods $\widehat S=\widehat S^{(i)}$, $i=0,\ldots,5$ the MAD on the log-transformed intensities is defined as:
 \begin{equation*}
\mathrm{MAD}(\log(\widehat S)) = \frac{1}{N} \sum _{\ell =1}^N \left( \frac{1}{n_{\text{reg}}} \sum _{j=1}^{n_{\text{reg}}} \left| \log(\widehat{S}(X_j^\ell|\widehat\Theta_\ell)) -\log( S_j^\ell) \right| \right)  
 \end{equation*}
For each method $\widehat S^{(i)}$, we compute the excess risk ratio:
\begin{equation*}
R(i) = \mathrm{MAD}(\widehat \log (S^{(i)}))/\mathrm{MAD}(\widehat \log(S^{(0)})), \quad \text{for } i=1,\ldots,5
\end{equation*} 
The results are displayed in Table 5. 

\subsection{Robustness with respect to non-normal background noise.}
 We check the robustness of the normal-gamma BgC with respect to heavier right tails in the negative probe distribution from the simulation data set $(S_2)$, and compare the results obtained with normexp-NP. Figure D presents the density of the mixture distribution used for checking robustness.  Table B presents the MAD computed from the normal-gamma BgC  (first column) and the normexp BgC with NP estimation (second column).

\begin{table}[!h]
\begin{center} \begin{tabular}{crr}\hline
$p$  &  $\mathrm{MAD}(\widehat S^{(1)})$ & $\mathrm{MAD}(\widehat S^{(4)})$   \\\hline
0 & 2.38 & 3.59 \\
0.10 & 3.12 & 4.88 \\
0.25 & 4.19 & 6.51 \\
0.50 & 5.70 & 8.72 \\
0.75 & 6.82 & 10.53 \\
1 & 7.59 & 12.01 \\\hline
\end{tabular} \end{center}
\caption*{Table B. MAD of background corrected intensities from simulated data with background distribution defined as a mixture $(1-p)\mathcal N(50,4)+p\,\upchi^2(3,55)$ and signal is generated from a $\Gamma (0.12,1785)$. }
\end{table}

\newpage

\begin{figure}[!h]
\begin{center}
  \includegraphics[height=14cm, width=18cm]{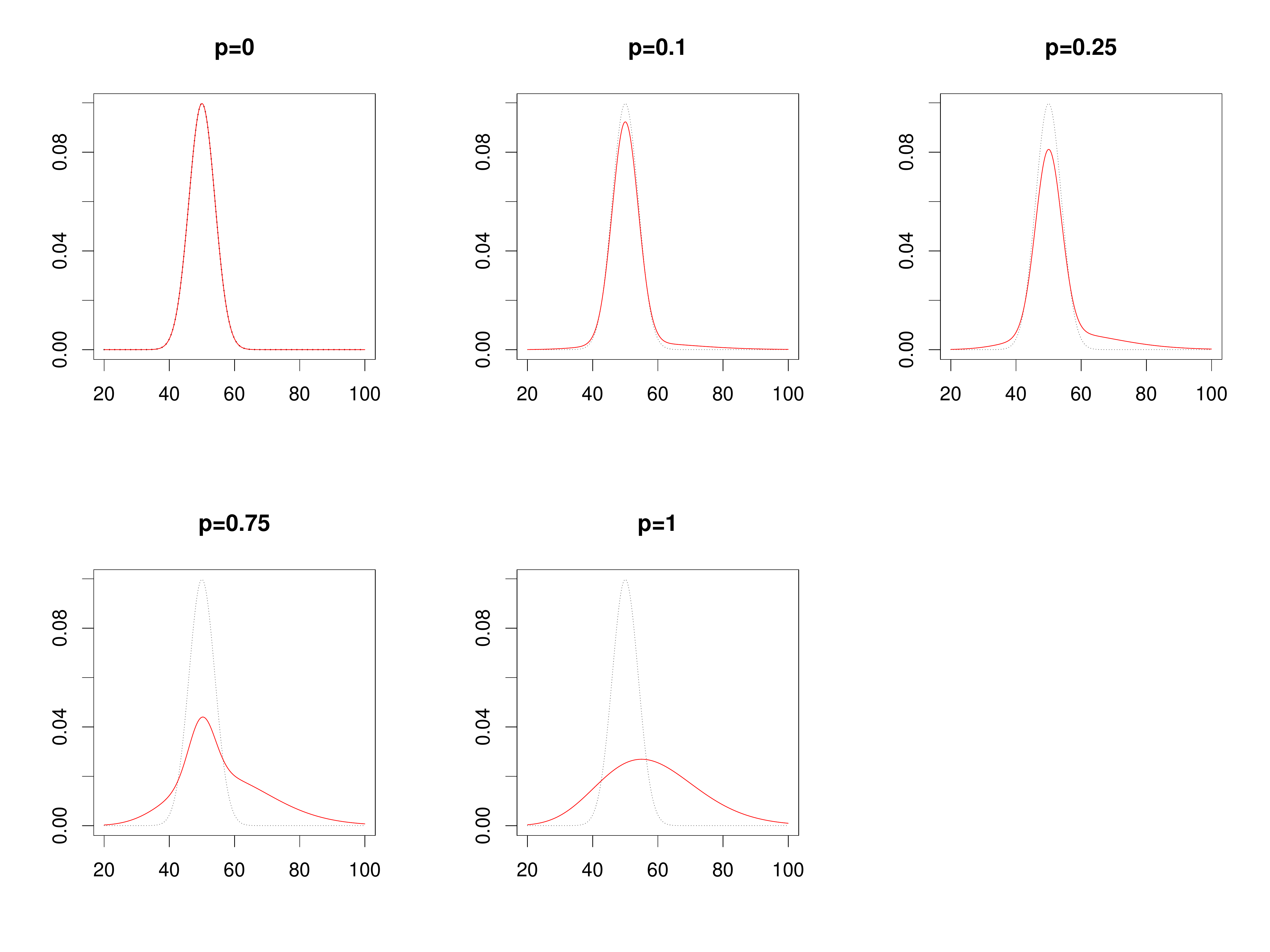}
  \caption*{Figure C. Density of the mixture distribution $(1-p)\mathcal N(53,4.4)+p\, \upchi^2(3,55)$ for various values of $p$.}
\end{center}
\end{figure}

\newpage

\subsection{Operating characteristics}

\subsubsection{Equal Innate offset on spike-in data set $(E_3)$}

An offset has been added to the regular probe intensities prior to log-transformation for each method, in order to equalize innate offsets. The results displayed in Figure D corresponds to a total offset equal to 100 (i.e. an added offset of 98.5 for the normal-gamma BgC, 76.5 for normexp-MLE, 93 for normexp-RMA  and 87.6 for normexp-NP).  The first column presents the average intensity, the second column displays the estimated log-ratio for each pair of consecutive concentrations, corresponding to a true fold-change of 3 or 3.33.. The thirs column shows the standart deviation between replicates. 

We observe that the precision is globally similar, but the difference on small concentration is enlighted by the observation of the fold-change for every pair of consecutive concentrations. We note that normexp-MLE provides a smaller bias (larger log-ratios) and a poorer precision.

\begin{figure}[!h]
\begin{center}
 \includegraphics[height=6cm, width=18cm]{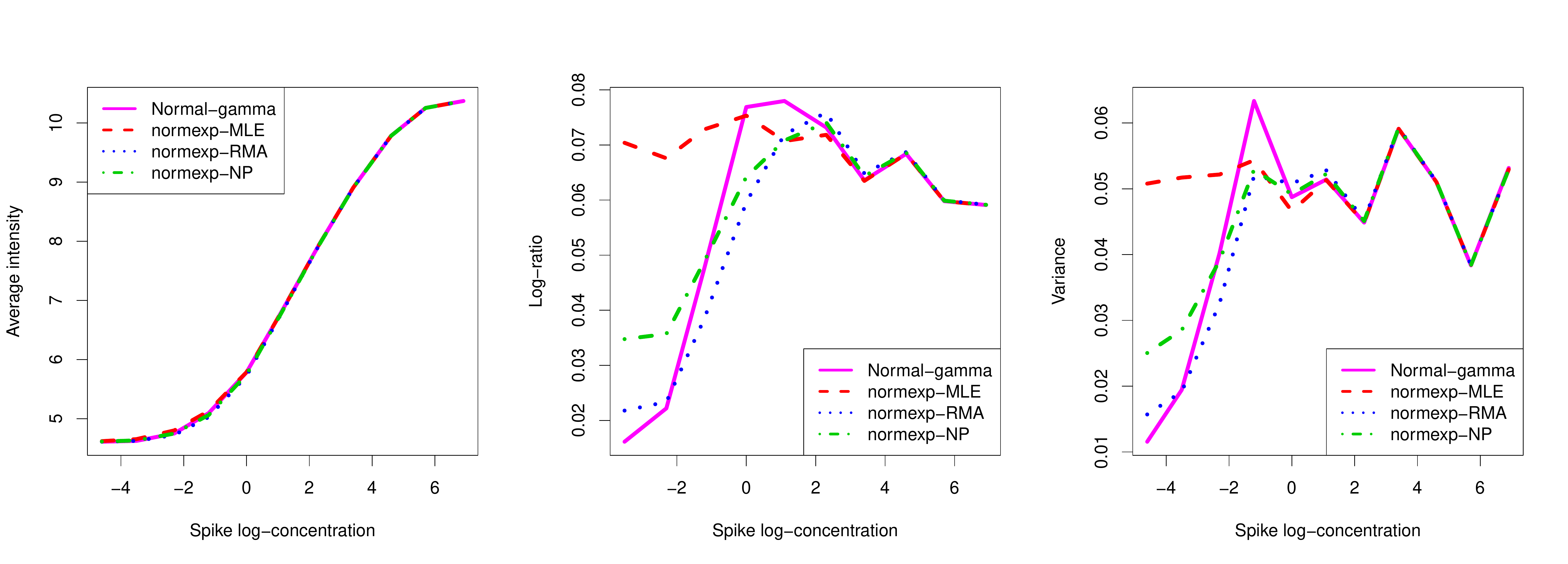} 
 \caption*{Figure D. Average spike intensity (left), average log-ratio for every pair of consecutive spike concentrations (center), and average standard deviation between spike replicates (right) as a function of the spike log-concentration after equalisation of the innate offsets on raw data. }
\end{center}
\end{figure}

\subsubsection{Operating characteristics on normal-exponential simulated data}

The operating characteristics of the four BgC methods on the data set $(S_3)$  with parameter set 8 are displayed in Figure F.

\begin{figure}[!h]
\begin{center}
\includegraphics[height=6cm, width=12cm]{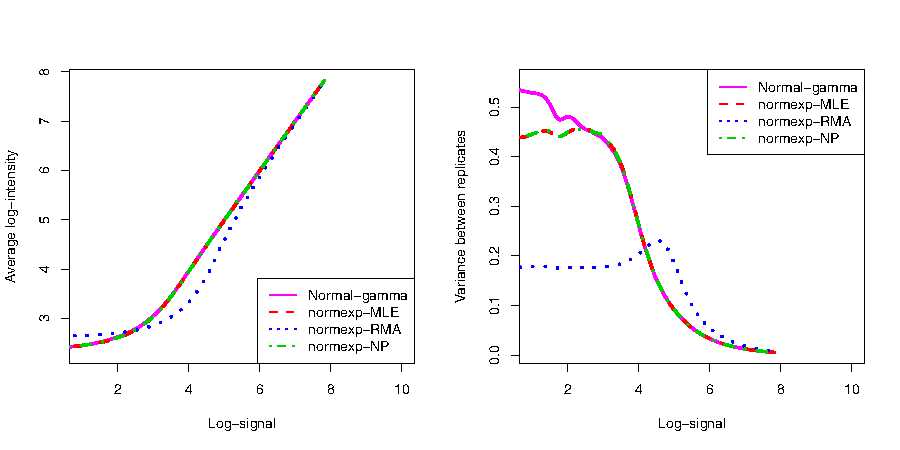} 
\caption*{Figure E. Average log-intensities (left) and log-variance between replicates (right) for simulation data set $(S_3)$  with parameter set 8.}
\end{center}
\end{figure}

\newpage

\subsection{Numerical implementation}

\subsubsection{Expression of the normal-gamma density}

The convolution product of the normal and gamma density:
\begin{equation*}
f^{\text{ng}}_{\mu, \sigma , k,\theta }  (x) = \int f^{\text{gam}}_{k,\theta } (t) f^{\text{norm}}_{\mu, \sigma } (x-t) dt
\end{equation*}
does not have an analytic expression similar to the normal-exponential density. Nevertheless, an expression involving the Kummer's special mathematical function may be obtained by using the software {\tt Mathematica}.  Denoted by kum the M-Kummer's special function,
 $$\begin{array}{l}
\hspace{-0.3cm}\vspace{0.2cm}f^{\text{ng}}_{\mu, \sigma , k,\theta }  (x)  = 2^{1+k/2} \left( \frac{\sigma }{\theta} \right)^k f^{\text{norm}} _{\mu,\sigma} (x) \cdot \\
\hspace{-0.3cm} \left[ \Gamma \left( \frac{k}{2} \right) \text{kum} \left( \frac{k}{2} , \frac{1}{2} , \frac{y^2}{2} \right) - \sqrt{2} y \Gamma \left( \frac{k+1}{2} \right) \text{kum} \left( \frac{k+1}{2} , \frac{3}{2} , \frac{y^2}{2} \right) \right] 
\end{array}$$
 where $y= (\sigma^2 + \theta (\mu -x) )/(\sigma \theta)$. The R-package {\tt fAsianOptions}  implements the M-Kummer's special function. Unfortunately, it does not offer enough numerical  stability to be used for  microarray expressions, due to their range.

\subsubsection{Computation of the normal-gamma density estimator by {\tt fft}.}

The normal gamma density $f_X = f_{\mu , \sigma , k ,\theta} ^{\text{ng}} $ is computed with the Fast Fourier Transform ({\tt fft}) function on a regular grid of $[0,T]$ where $T= \mu+5\sigma+ q$ and $q$ is the gamma distribution 0.99999-quantile, then interpolated on other points. \\

\noindent \textbf{Approximation of $f_X$.} 
Let $\varphi _X $ denote the moment generating function of $X$,
$$\varphi _X (t) = \mathbb{E} \left[ e^{itX} \right] , \quad \forall t \in \mathbb{R}.$$
Then $$\varphi _X (t) = \varphi _S (t) \varphi _B (t)= (1-it\theta) ^{-k} e^{i\mu t} e^{-\sigma^2 t^2 /2} .$$
Moreover, for every $x \in \mathbb{R}$,

\begin{eqnarray*}
f_X(x) & = & \int _{-\infty}^{\infty} \varphi _X (t)e^{-itx} dt \\
 & \simeq &  \int _{-A}^A \varphi _X (t)e^{-itx} dt \\
 & \simeq &  \frac{2A}{N} \sum _{j=0}^{N-1}  \varphi _X \left(- A + \frac{2A}{N} (j-1)\right) \\
 &&  \exp \left( -itx \left(- A + \frac{2A}{N} (j-1)\right) \right)   \quad \text{(Riemann sum)}
\end{eqnarray*}
for large $A$ and $N$ and small ratio $A/N$. Let us denote by $\overline{f} _X$ this approximate. \\

\noindent \textbf{Computation using {\tt fft}} 

The {\tt fft} function of {\tt R} is defined as follows. For a vector ${\bf V} $ of length $N$,
$${\tt fft } ({\bf V}) = \sum _{j=1}^N {\bf V} [j] \exp \left( -\frac{2i\pi }{N} (j-1)(k-1) \right).$$
Let 
\begin{eqnarray*}
&& {\bf U } = \frac{\pi}{A} (0:N-1) \\
&& {\bf V} = \varphi _X \left( -A + \frac{2A}{N} (0:N-1) \right) 
\end{eqnarray*}
and 
$$ {\bf W} = \frac{A}{N\pi} \exp ( i A {\bf U}) {\tt fft} ( {\bf V})$$ 
then  for $k=0, \dots , N-1$, 

\begin{eqnarray*}
{\bf W } [k] &=&  \frac{A}{N\pi} \exp (i\pi (k-1) ) \sum _{j=1}^N \varphi _X \left( -A + \frac{2A}{N} (j-1) \right) \\
&& \exp \left(-\frac{2i\pi}{N} (j-1)(k-1) \right) = \overline{f}_X(\frac{k-1}{A}).
\end{eqnarray*}

\subsubsection{Normal-gamma MLE computation.}

The parameters $(\mu,\sigma,k,\theta)$ of the normal-gamma model are estimated by likelihood maximization, ensured by the R-function {\tt optimx} with the following initialization values.
\begin{eqnarray*} 
\mu_0 &=& {\rm mean}(X_j,j\in J_0),\\
\sigma_0 &=& IQR(X_j,j\in J_0)/1.349,\\
\theta_0 &=& (\text{sd}(X_j,j\in J)^2-\sigma_0^2)/({\rm mean}(X_j,j\in J)-\mu_0),\\
k_0 &=& ({\rm mean}(X_j,j\in J)-\mu_0)/\theta_0,
\end{eqnarray*}
where $IQR$ designs the interquartile range and the new parametrization
\[ (p_1,p_2,p_3,p_4) = (\mu, \sigma, k\theta, \theta\sqrt k), \]
in order to get homogenous parameters.

\newpage

\subsection{Infer the negative probe intensities from Illumina detection p-values}

\subsubsection{Algorithm}

For a given array, let $X$ and $N$ be the vectors of regular and negative probe intensities, and $P$ the vector of detection p-values. For each regular probe $j$ with intensity $X_j$, the detection p-value $P_j$ is equal to the proportion of negative probes which intensities are larger than $X_j$:
$$P_j = \frac{1}{n_{\text{neg}}} Card \left\{  k,  N_ k > X_j \right\}$$
where $n_{\text{neg}}$ is the number of negative probes.  Let $Q$ be the vector of ordered unique values of $P$, and denote by $\ell (Q)$ its length. For every $k \in \{ 2, \dots, \ell (Q)\}$, let:

$$x_k^ 1 = \max \{ X_j, P_j =  Q_k \} $$
$$x_k^2 = \min \{ X_j, P_j = Q_{k+1} \}$$

then 
$$Q_{k+1} - Q_k = \frac{k}{n_{\text{neg}}} $$
where $d$ is the number of negative probes whose intensity lies in $[x_k^1, x_k^2]$. Therefore, for every $k \in \{ 2, \dots, \ell (Q)\}$, we infer $d$ negative probes with intensity equal to 
$$\frac{1}{2} (x_k^1 + x_k^2 ) .$$

Moreover, if $\min(Q)=0$, we infer a negative probe with intensity 
$  \max \{ X_j, P_j =  0 \}$
and if  $\min(Q)=d_0/ n_{\text{neg}}$, we infer $d_0$ probes with intensity equal to $\min(X)$. \\


\subsubsection{Performances}

For the ten human microarrays, we have computed the parameters 1) with the true negative probe intensities, 2) with the set of inferred negative probe intensities, for the normal-gamma model and for the normexp model with MLE and NP parameters (the RMA algorithm do  not uses the negative probe intensities).  Table C presents the relative error on parameters and on background corrected intensities.

\begin{table}[!h]
\begin{center}
\begin{tabular}{cccccc}\hline 
\rule{0pt}{15pt}\vspace{0.2cm}Method & $\mu$ & $\sigma $ & $k$ & $\theta$ & $\widehat{S}$ \\\hline 
\rule{0pt}{15pt} \vspace{0.2cm}Normal-gamma & 2.9 E-5 & 5.2  E-4  &7.9 E-4 &  4.2 E-4 & 6.3 E-4  \\
\rule{0pt}{15pt}\vspace{0.2cm}normexp-MLE & 4.7 E-6 & 5.8 E-5 & $\star $ & 7.2 E-4 & 2.1 E-5  \\
\rule{0pt}{15pt}\vspace{0.2cm}normexp-NP & 1.4 E-5& 1.8 E-4 & $\star$ & 7.2 E-6 & 1.3 E-4  \\\hline 
 \end{tabular}\end{center}
\caption*{Table C. Relative error of estimation from replacing negative probe intensities by inferred values. Column 2-5: parameters of the model; Column 6: background corrected values. }
 \end{table}

\section*{Author's contributions}

    {\bf Grant:} ERC-2008-AdG 232997-TICE "Transcriptomics in cancer epidemiology".
    
\noindent {\bf Contribution:}The NOWAC data were provided by Eiliv Lund, Principal Investigator of TICE project. 
Statistical and computational aspects were developed by Sandra Plancade and Yves Rozenholc.

\section*{Acknowledgements}
  The authors thank Gregory Nuel for fruitful discussions on numerical issue. 
\newpage

 \bibliographystyle{plain} 
  \bibliography{Biblio-Nowac}   

\begin{thebibliography}{10}

\bibitem{Shi-maqc}
The microarray quality control (maqc) project shows inter- and intraplatform
  reproducibility of gene expression measurements.
\newblock {\em Nat Biotechnol}, 25(9):11--51, 2006.

\bibitem{Ding08}
Liang-Hao Ding, Yang Xie, Seongmi Park, Guanghua Xiao, and Michael~D Story.
\newblock Enhanced identification and biological validation of differential
  gene expression via illumina whole-genome expression arrays through the use
  of the model-based background correction methodology.
\newblock {\em Nucleic Acids Res}, 36(10):e58, Jun 2008.

\bibitem{Dunning08}
Mark~J Dunning, Nuno~L Barbosa-Morais, Andy~G Lynch, Simon Tavar{\'e}, and
  Matthew~E Ritchie.
\newblock Statistical issues in the analysis of illumina data.
\newblock {\em BMC Bioinformatics}, 9:85, 2008.

\bibitem{Gleser89}
Leon~Jay Gleser.
\newblock The gamma distribution as a mixture of exponential distributions.
\newblock {\em Amer. Statist.}, 43(2):115--117, 1989.

\bibitem{Irizarry03}
Rafael~A Irizarry, Bridget Hobbs, Francois Collin, Yasmin~D Beazer-Barclay,
  Kristen~J Antonellis, Uwe Scherf, and Terence~P Speed.
\newblock Exploration, normalization, and summaries of high density
  oligonucleotide array probe level data.
\newblock {\em Biostatistics}, 4(2):249--64, Apr 2003.

\bibitem{Langaas05}
Mette Langaas, Bo~Henry Lindqvist, and Egil Ferkingstad.
\newblock Estimating the proportion of true null hypotheses, with application
  to dna microarray data.
\newblock {\em Journal of the Royal Statistical Society Series B},
  67(4):555--572, 2005.

\bibitem{Lin08}
Simon~M Lin, Pan Du, Wolfgang Huber, and Warren~A Kibbe.
\newblock Model-based variance-stabilizing transformation for illumina
  microarray data.
\newblock {\em Nucleic Acids Res}, 36(2):e11, Feb 2008.

\bibitem{Lund-nowac}
Eiliv Lund, Vanessa Dumeaux, Tonje Braaten, Anette Hjart{\aa}ker, Dagrun
  Engeset, Guri Skeie, and Merethe Kumle.
\newblock Cohort profile: The norwegian women and cancer study--nowac--kvinner
  og kreft.
\newblock {\em Int J Epidemiol}, 37(1):36--41, Feb 2008.

\bibitem{Lynch10}
Andy~G. Lynch, James Hadfield, Mark~J. Dunning, Michelle Osborne, Natalie~P.
  Thorne, and Simon Tavar\'{e}.
\newblock The cost of reducing starting rna quantity for illumina beadarrays: a
  bead-level dilution experiment.
\newblock {\em BMC genomics}, 11, 2010.

\bibitem{Irizarry08}
Matthew N~N. McCall and Rafael A~A. Irizarry.
\newblock Consolidated strategy for the analysis of microarray spike-in data.
\newblock {\em Nucleic Acids res.}, 3:e108, 2008.

\bibitem{Ritchie07}
Matthew~E Ritchie, Jeremy Silver, Alicia Oshlack, Melissa Holmes, Dileepa
  Diyagama, Andrew Holloway, and Gordon~K Smyth.
\newblock A comparison of background correction methods for two-colour
  microarrays.
\newblock {\em Bioinformatics}, 23(20):2700--7, Oct 2007.

\bibitem{Rocke01}
D~M Rocke and B~Durbin.
\newblock A model for measurement error for gene expression arrays.
\newblock {\em J Comput Biol}, 8(6):557--69, 2001.

\bibitem{rozenholc-thoralf}
Y.~Rozenholc, T.~Mildenberger, and U.~Gather.
\newblock Constructing irregular histograms by penalized likelihood.
\newblock {\em Computational Statistics and Data Analysis}, 54(12):3313--23,
  2010.

\bibitem{Shi10b}
Wei Shi, Carolyn~A. de~Graaf, Sarah~A. Kinkel, Ariel~H. Achtman, Tracey
  Baldwin, Louis Schofield, Hamish~S. Scott, Douglas~J. Hilton, and Gordon~K.
  Smyth.
\newblock {Estimating the proportion of microarray probes expressed in an RNA
  sample}.
\newblock {\em Nucleic Acids Research}, 38(7):2168--2176, 2010.

\bibitem{Shi10}
Wei Shi, Alicia Oshlack, and Gordon~K Smyth.
\newblock Optimizing the noise versus bias trade-off for illumina whole genome
  expression beadchips.
\newblock {\em Nucleic Acids Res}, 38(22):e204, Dec 2010.

\bibitem{Silver08}
Jeremy~D Silver, Matthew~E Ritchie, and Gordon~K Smyth.
\newblock Microarray background correction: maximum likelihood estimation for
  the normal-exponential convolution.
\newblock {\em Biostatistics}, 10(2):352--63, Apr 2009.

\bibitem{Smyth05}
G.K. Smyth.
\newblock Limma: linear models for microarray data.
\newblock In {\em Bioinformatics and Computational Biology Solutions using R
  and Bioconductor}, pages 397--420. In Gentleman,R. et al. (eds.), Springer,
  New York, 2005.

\bibitem{Smyth04}
Gordon~K Smyth.
\newblock Linear models and empirical bayes methods for assessing differential
  expression in microarray experiments.
\newblock {\em Stat. Appl. Gen. Mol. Biol.}, 3(1), 2004.

\bibitem{Wang11}
Xiao-Feng Wang and Deping Ye.
\newblock The effects of error magnitude and bandwidth selection for
  deconvolution with unknown error distribution.
\newblock {\em Journal of Nonparametric Statistics}, 24(1):153--167, 2012.

\bibitem{Xie09}
Yang Xie, Xinlei Wang, and Michael Story.
\newblock Statistical methods of background correction for illumina beadarray
  data.
\newblock {\em Bioinformatics}, 25(6):751--7, Mar 2009.

\end{thebibliography}

\end{document}